\documentclass[preprint,showpacs,preprintnumbers,nofootinbib,amsmath,amssymb]{revtex4}
\usepackage{graphicx}
\usepackage{dcolumn}
\usepackage{bm}
\begin{document}
\title{Pairing in hot
rotating nuclei}
 \author{N. Quang Hung$^{1}$}
 \altaffiliation[On leave of absence from the ]
 {Institute of Physics and Electronics, Hanoi, Vietnam}
  \email{nqhung@riken.jp}
  \author{N. Dinh Dang$^{1, 2}$}
 \email{dang@riken.jp}
\affiliation{1) Heavy-Ion Nuclear Physics Laboratory, RIKEN Nishina Center
for Accelerator-Based Science,
2-1 Hirosawa, Wako City, 351-0198 Saitama, Japan\\
2) Institute for Nuclear Science and Technique, Hanoi, Vietnam}
\date{\today}
\begin{abstract}
Nuclear pairing properties are studied within an approach that
includes the quasiparticle-number fluctuation (QNF) and coupling to
the quasiparticle-pair vibrations at finite temperature and angular
momentum. The formalism is developed to describe non-collective
rotations about the symmetry axis. The numerical calculations are
performed within a doubly-folded equidistant multilevel model as
well as several realistic nuclei. The results obtained for the
pairing gap, total energy and heat capacity show that the QNF
smoothes out the sharp SN phase transition and leads to the
appearance of a thermally assisted pairing gap in rotating nuclei at
finite temperature. The corrections due to the dynamic coupling to
SCQRPA vibrations and particle-number projection are analyzed. The
effect of backbending of the momentum of inertia as a function of
squared angular velocity is also discussed.
\end{abstract}

\pacs{21.60.Jz, 21.60.-n, 24.10.Pa, 24.60.-k}
\keywords{Suggested keywords}
\maketitle

\section{INTRODUCTION}
\label{sec1}
Thermal effect on pairing correlations
has been extensively studied within the
Bardeen-Cooper-Schrieffer (BCS) theory \cite{BCS} at finite
temperature $T$ (FTBCS theory). The FTBCS theory
predicts a destruction of pairing correlation at a critical
temperature $T_{\rm c}\simeq 0.568\Delta(0)$ [$\Delta(0)$ is the
pairing gap at zero temperature],  resulting in a sharp transition from the
superfluid phase to normal one (the SN phase transition)
in good agreement with the experimental findings in
macroscopic systems such as metallic
superconductors.
However, the BCS theory is valid only when the assumption on the
quasiparticle mean field is good, i.e. when the difference between the
pair correlator $P_{k\sigma}^{\dagger}\equiv
a_{k\sigma}^{\dagger}a_{k-\sigma}^{\dagger}$
and its expectation value
$\langle P_{k\sigma}^{\dagger}\rangle$ is small
so that the quadratic term
$(P_{k\sigma}^{\dagger}-
\langle P_{k\sigma}^{\dagger}\rangle)^{2}$ is
negligible, where $a_{k\sigma}^{\dagger}$ is the operator that creates a particle
with angular momentum $k$ and spin $\sigma$. For small systems such as
underdoped cuprates, where
the coherence lengths (the Cooper-pair sizes) are very short, the
fluctuations $(P_{k\sigma}^{\dagger}-
\langle P_{k\sigma}^{\dagger}\rangle)^{2}$ are
no longer small, which invalidate the quasiparticle mean-field
assumption, and break down the BCS theory.
As the result, the gap evolves
continuously across $T_{\rm c}$, and persists well
above $T_{\rm c}$~\cite{Ding}.

Various theoretical studies have been undertaken in the last three
decades to study the effects of
thermal fluctuations on pairing in atomic nuclei. Pioneer papers by
Moretto~\cite{MorettoPLB40} employed
the macroscopic Landau theory of phase
transition to treat thermal fluctuations in the pairing
field as those occurring around the
most probable value of the pairing gap. The
results of calculations within the uniform model carried out in Ref.
\cite{MorettoPLB40} show that
the average pairing gap does not collapse as predicted
by the FTBCS theory, but decreases monotonously with increasing $T$,
smearing out the sharp SN phase transition.
This approach was later used by Goodman
to include the effects of thermal fluctuations
in the Hartree-Fock-Bogoliubov (HFB) theory
at finite temperature
\cite{GoodmanPRC29}. The static-path
approximation (SPA), which
takes into account thermal fluctuations by averaging
over all static paths around the mean field,
also shows the non-collapsing pairing gap at
finite temperature \cite{DangPLB297,DangPRC47}.
The recent microscopic approach called the modified BCS (MBCS) theory
\cite{DangPRC64,DangPRC67,DangNPA784} is based on the secondary
Bogoliubov transformation from quasiparticles to the modified ones
to restore the unitary relation for the generalized particle-density
matrix at $T\neq$ 0. The MBCS theory, for the first time, points
out the quasiparticle-number fluctuation (QNF) as
the microscopic origin that causes
the non-collapsing thermal pairing gap in finite small systems. The
predictions of these approaches are in qualitative agreements with
the experimental findings of pairing gaps and heat capacities measured in underdoped
cuprates~\cite{Ding} and extracted from nuclear level
densities~\cite{Schiller}.

While quasiparticles are regarded as independent in
all above-mentioned approaches, the recently proposed
FTBCS1 theory with corrections coming from the QNF and
self-consistent quasiparticle random-phase approximation (SCQRPA)
at finite temperature~\cite{DangHungPRC2008} calculates the
quasiparticle occupation numbers from a set of FTBCS1+SCQRPA equations.
Within this approach, which is called the FTBCS1+SCQRPA and is an
extension of the BCS1+SCQRPA developed in Ref.
\cite{HungDangPRC76} to finite temperature, the QNF and quantal fluctuations caused by
coupling to
SCQRPA vibrations are included into the equations for the pairing gap and
particle number. Under the influence of these SCQRPA corrections, the
temperature functional of the quasiparticle occupation number deviates
from the Fermi-Dirac
distribution of independent quasiparticles.
The results obtained within the FTBCS1+SCQRPA for the total
energies
and heat capacities
agree fairly well with the exact solutions of the
Richardson model~\cite{Richardson,Volya} at finite temperature, and those obtained within the
finite-temperature quantum Monte Carlo method for the realistic
$^{56}$Fe nucleus~\cite{QMC}.

The positive results of the FTBCS1+SCQRPA encourage a further extension of
this approach to include the effect of angular momentum on nuclear
pairing so that it can be applied to study hot rotating nuclei.
The rotational phase of nucleus as a whole,
such as that present in spherical nuclei,
or that about the axis of symmetry in deformed nuclei,
is
known to affect nuclear level densities.
The relationship between this noncollective rotation
and pairing correlations has been the subjects of many theoretical
studies. The effect of thermal pairing on the angular momentum at
finite temperature was first examined
by Kammuri in Ref.~\cite{Kammuri}, who included in
the FTBCS equations the effect caused by the
projection $M$ of the total angular momentum operator on the $z$-axis
of the laboratory system (or nuclear symmetry axis in the case of
deformed nuclei).
It has been pointed out in
Ref.~\cite{Kammuri} that, at finite angular
momentum, a system can turn into the superconducting phase at some
intermediate excitation energy (temperature), whereas it remains in
the normal phase at low and high excitation energies. This
effect was later confirmed by Moretto in Refs.~\cite{MorettoPLB35,MorettoNPA185}
by applying the FTBCS at finite angular momentum to
the uniform model. It has been shown in these papers that,
apart from the region where the pairing
gap decreases with increasing both temperature $T$ and angular
momentum $M$, and vanishes at a given critical values $T_{\rm c}$ and
$M_{\rm c}$, there is a region of $M$, whose
values are slightly higher than
$M_{\rm c}$, where
the pairing gap reappears at $T = T_{1}$, increases with $T$ at
$T> T_{1}$ to reach a maximum, then decreases again to vanish at $T \geq
T_{2}$. This effect is called anomalous pairing or thermally
assisted pairing correlation. In the recent study of the projected gaps
for even or odd number of particles in ultra-small metallic grains in
Ref. \cite{BalianPR317} a similar reappearance of pairing correlation at finite
temperature was also found, which is referred to as the reentrance
effect. Recently, this phenomenon was further studied
in Refs. \cite{FrauendorfPRB68,SheikhPRC72}
by performing the calculations using the exact pairing
eigenvalues embedded in the canonical ensemble
at finite temperature and rotational frequency.
The results of Refs. \cite{FrauendorfPRB68,SheikhPRC72}
also show a manifestation of the reentrance of pairing
correlation at finite temperature. However,  different from the
results of the FTBCS theory,
the reentrance effect shows up in such a way that the pairing
gap reappears at a given $T = T_1$ and remains finite at $T
> T_1$ due to the strong fluctuations of the order
parameters.

The aim of the present study is to extend the FTBCS1 (FTBCS1+SCQRPA)
theory of Ref. \cite{DangHungPRC2008}
to finite angular momentum so that both the effects of angular momentum
as well as QNF on nuclear pairing correlation can be studied
simultaneously in a microscopic way.
The formalism is applied to a doubly degenerate equidistant model
with a constant pairing interaction $G$ and some realistic nuclei,
namely $^{20}$O, $^{22}$Ne, and $^{44}$Ca.

The paper is organized as follows. The
FTBCS1+SCQRPA theory is extended to include
a specified
projection $M$ of the total angular momentum on the axis of quantization
in
Sec. \ref{sec2}. The results of numerical
calculations are discussed in Sec. \ref{sec3}. The last
section summarizes the paper, where conclusions are drawn.

\section{\label{sec2}FORMALISM}

\subsection{\label{subsec1}Model Hamiltonian}
We consider a system of $N$ particles interacting via a pairing
force with the parameter $G$, and rotating about the
symmetry axis (noncollective rotation) at an
angular velocity (rotational frequency) $\gamma$ with a fixed
projection $M$ (or $K$) of the total angular momentum operator
along this axis. For a spherically symmetric system,
it is always possible to make the laboratory-frame $z$ axis,
taken as the axis of quantization, coincide with the body-fixed
one, which is aligned within the quantum mechanical uncertainty with the
direction of the total angular momentum, so that the latter
is completely determined by its $z$-axis projection
$M$ alone. As for deformed systems,
where the axis symmetry is the principal (body-fixed) axis,
this noncollective motion is known as the
``single-particle'' rotation, which takes place
when the angular momenta of individual nucleons are aligned parallel to
the symmetry axis, resulting in an axially symmetric oblate shape
rotating about this axis. Such noncollective motion is also possible
in high-$K$ isomers~\cite{BM}, which have many single-particle orbitals near the
Fermi surface with a large and approximately conserved projection $K$
of individual nucleonic
angular momenta along the symmetry axis.
Therefore, without losing generality, further derivations are carried
out below for the pairing Hamiltonian of a
spherical system rotating about the
$z$ axis~\cite{Kammuri,MorettoPLB35,MorettoNPA185}, namely
\begin{equation}\label{Ha}
    H=H_{P}- \lambda\hat{N} - \gamma\hat{M}~,
    \end{equation}
    where $H_{P}$ is the well-known pairing Hamiltonian
    \begin{equation}\label{Ha1}
          H_{P}=\sum_{k}\epsilon_{k}(N_{k}+N_{-k})
-G\sum_{k,k'}P_{k}^{\dagger}P_{k'}~,
        \hspace{5mm} N_{\pm
        k}=a_{\pm k}^{\dagger}a_{\pm k}~,\hspace{5mm}
        P_{k}^{\dagger}=a_{k}^{\dagger}a_{-k}^{\dagger}~,
    \end{equation}
with $a_{\pm k}^{\dagger}$ ($a_{\pm k}$) denoting the operator that
creates (annihilates) a particle with angular momentum $k$, spin
projection $\pm m_{k}$, and energy $\epsilon_k$.
For simplicity, the subscripts $k$
are used to label the single-particle states
$|k,m_{k}\rangle$ in the deformed basis with
the positive single-particle spin
projections $m_{k}$, whereas the subscripts $-k$  denote the
time-reversal states $|k,-m_{k}\rangle$ ($m_{k}>$ 0).
The particle number operator $\hat{N}$
and angular momentum $\hat{M}$ can be expressed in terms of
a summation over the single-particle levels:
    \begin{equation}\label{N&M}
        \hat{N}=\sum_k(N_k+N_{-k})~,\hspace{5mm}
        \hat{M}=\sum_{k}m_{k} (N_{k}-N_{-k})~,
    \end{equation}
whereas the chemical potential $\lambda$ and angular velocity
$\gamma$ are two Lagrangian multipliers to be determined.
For deformed and axially symmetric
systems, the $z$-projection $M$ and spin projections $m_{k}$ should be
identified with the projection $K$ along the
body-fixed symmetry axis and spin projections
$\Omega_{k} $, respectively, which are good quantum numbers~\cite{MorettoNPA185}.

By using the Bogoliubov transformation from particle operators,
$a_k^{\dagger}$ and $a_k$, to quasiparticle ones, $\alpha_k^{\dagger}$
and $\alpha_k$,
    \begin{equation}\label{ak}
        a_k^{\dagger}=u_k\alpha_k^{\dagger} + v_k\alpha_{-k}~, \hspace{5mm}
        a_{-k}=u_k\alpha_{-k}-v_k\alpha_{k}^{\dagger}~,
    \end{equation}
the Hamiltonian \eqref{Ha} is transformed into the
quasiparticle Hamiltonian as
    \[
        \mathcal{H}=a+\sum_k{b_k^{+}\mathcal{N}_k^{+}}+\sum_{-k}{b_k^-\mathcal{N}_k^{-}}
        +\sum_k{c_k(\mathcal{A}_k^{\dagger}+\mathcal{A}_k})
        +\sum_{kk'}{d_{kk'}\mathcal{A}_k^{\dagger}\mathcal{A}_{k'}}
        +\sum_{kk'}{g_k(k')(\mathcal{A}_{k'}^{\dagger}\mathcal{N}_k+\mathcal{N}_k\mathcal{A}_{k'})}
    \]
    \begin{equation}\label{QHa}
        +\sum_{kk'}{h_{kk'}(\mathcal{A}_k^{\dagger}\mathcal{A}_{k'}^{\dagger}
        +\mathcal{A}_{k'}\mathcal{A}_k)}+\sum_{kk'}{q_{kk'}\mathcal{N}_k\mathcal{N}_{k'}}~,
    \end{equation}
where $\mathcal{N}_k^{+}$ and $\mathcal{N}_k^{-}$ are the
quasiparticle-number operators, whereas $\mathcal{A}_k^{\dagger}$ and
$\mathcal{A}_k$ are the creation and destruction operators of a pair
of time-conjugated quasiparticles, respectively:
    \begin{eqnarray}
        &&\mathcal{N}_k^{+}=\alpha_k^{\dagger}\alpha_k ~, \hspace{5mm}
        \mathcal{N}_k^{-}=\alpha_{-k}^{\dagger}\alpha_{-k} ~,
        \hspace{5mm}
        \mathcal{N}_k=\mathcal{N}_k^{+} + \mathcal{N}_k^{-} \label{Nk}~,\\
        &&\mathcal{A}_k^{\dagger}=\alpha_k^{\dagger}\alpha_{-k}^{\dagger}~,
        \hspace{5mm}
        \mathcal{A}_k=(\mathcal{A}_k^{\dagger})^{\dagger}~.\label{Ak}
    \end{eqnarray}
They obey the following commutation relations
    \begin{eqnarray}
        &&[{\cal A}_k~,~{\cal A}_{k'}^{\dagger}]=\delta_{kk'}{\cal D}_k ~,\hspace{2mm}{\rm where}
        \hspace{2mm} {\cal D}_{k}=1-{\cal N}_k ~,
        \label{[AA]}\\
        &&[{\cal N}^{\pm}_k ~,~ {\cal A}_{k'}^{\dagger}]=
        \delta_{kk'}{\cal A}_{k'}^{\dagger}~,
        \hspace{5mm}
        [{\cal N}^{\pm}_k ~,~ {\cal A}_{k'}]=-\delta_{kk'}{\cal A}_{k'}~.
        \label{[NA]}
    \end{eqnarray}
The coefficients $b_k^{\pm}$ in Eq. \eqref{QHa} are
given as
    \begin{equation}
        b_k^{\pm}\equiv b_{k}\mp \gamma
        m_k = (\epsilon_k-\lambda)(u_k^2-v_k^2)
        +2Gu_kv_k\sum_{k'}{u_{k'}v_{k'}}+Gv_k^4\mp\gamma
        m_k~,\label{bk}
    \end{equation}
whereas the expressions for the
other coefficients $a$, $b_{k}$, $c_k$, $d_{kk'}$, $g_k(k')$, $h_{kk'}$, and
$q_{kk'}$ in Eqs. \eqref{QHa} and (\ref{bk}) can be found, e.g., in
Refs.
\cite{HungDangPRC76,HogaasenNP28,DangZPA335}.
\subsection{\label{Gapeq}Gap and number equations}
We use the exact commutation relations \eqref{[AA]} and \eqref{[NA]},
and follow the same procedure
introduced in Ref. \cite{DangHungPRC2008}, which is based on
the variational method
    \begin{equation}
        \frac{\partial \langle\mathcal{H}\rangle}{\partial u_k} +
        \frac{\partial \langle\mathcal{H}\rangle}{\partial v_k}
        \frac{\partial v_k}{\partial u_k} \equiv
        \langle[\mathcal{H},\mathcal{A}_k^+]\rangle = 0 ~, \label{variation}
    \end{equation}
    to minimize the expectation value $\mathcal{H}$
    of the pairing Hamiltonian \eqref{QHa} in the grand canonical
    ensemble,
    \begin{equation}
        \langle\hat{\cal O}\rangle\equiv\frac{{\rm Tr} [\hat{\cal
        O}e^{-\beta {\cal H}}]}{{\rm Tr}e^{-\beta{\cal H}}}~,\label{average}
    \end{equation}
with $\langle\hat{\cal O}\rangle$ denoting the ensemble or thermal
average of the operator $\hat{\cal O}$.
The following gap equation is obtained, which formally looks like the
one derived in Refs. \cite{HungDangPRC76,DangHungPRC2008}, namely
    \begin{equation}
       \Delta_{k}=\frac{G}{\langle{\cal D}_{k}\rangle}
       {\sum_{k'}\langle{\cal D}_{k}{\cal D}_{k'}\rangle}
       u_{k'}v_{k'}~.
       \label{gapk}
    \end{equation}
Here
    \begin{equation}
        u_{k}^{2}=\frac{1}{2}\bigg(1
        +\frac{\epsilon'_{k}-Gv_{k}^{2}-\lambda}{E_{k}}\bigg)~,
        \hspace{5mm}
        v_{k}^{2}=\frac{1}{2}\bigg(1
        -\frac{\epsilon'_{k}-Gv_{k}^{2}-\lambda}{E_{k}}\bigg)~,
        \label{uv}
        \end{equation}
with the quasiparticle energies $E_{k}$ defined as
    \begin{equation}
        E_{k}=\sqrt{(\epsilon'_{k}-Gv_{k}^{2}
        -\lambda)^{2}+\Delta_{k}^{2}}~,
        \label{Ek}
    \end{equation}
where $\epsilon'_{k}$ are the
renormalized single particle energies:
    \begin{equation}
       \epsilon_{k}'=\epsilon_{k}+\frac{G}{\langle{\cal D}_{k}\rangle}
       \sum_{k'}(u_{k'}^{2}-v_{k'}^{2})\bigg
       (\langle{\cal A}_{k}^{\dagger}{\cal
       A}_{k'\neq k}^{\dagger}\rangle
       +\langle{\cal A}_{k}^{\dagger}{\cal A}_{k'}\rangle\bigg)~.
       \label{respe}
       \end{equation}
       Notice that the diagonal elements $\langle{\cal A}_{k}^{\dagger}{\cal
       A}_{k}^{\dagger}\rangle$ are excluded from all calculations
       because of the Pauli's principle.
The Bogoliubov's coefficients, $u_{k}$ and $v_{k}$, in Eq. (\ref{uv}) as well as the
quasiparticle energy $E_{k}$ in Eq. (\ref{Ek})
contain the self-energy correction $-Gv_{k}^{2}$. It describes the
change of the single-particle
energy $\epsilon_{k}$ as a function of the particle number starting
from the constant Hartree-Fock
single-particle energy as determined for a doubly-closed shell nucleus. This
self-energy correction is usually discarded in many
nuclear structure calculations, where experimental values or those obtained
within a phenomenological potential such as the Woods-Saxon one are
used for single-particle energies, on the ground that such self-energy correction
is already taken care of in the experimental or phenomenological
single-particle spectra.  As all calculations in the present paper
use the constant single-particle levels, determined at $T=$ 0 within
the schematic doubly-folded multilevel equidistant model and within the
Woods-Saxon potentials, we also choose to neglect, for simplicity, the self-energy
correction $-Gv_{k}^{2}$ from the right-hand sides of Eqs. (\ref{uv})
and (\ref{Ek}) in the numerical calculations.

       The expectation values
       $\langle{\cal A}_{k}^{\dagger}{\cal A}_{k'}^{\dagger}\rangle$
       and $\langle{\cal A}_{k}^{\dagger}{\cal A}_{k'}\rangle$ in Eq.
       (\ref{respe}) are
       called the screening factors. They are calculated by
       coupling to the SCQRPA vibrations in the next section.
The quasiparticle-number fluctuation (QNF) is included into the gap
equation following the exact treatment:
    \begin{equation}
        \langle{\cal D}_{k}{\cal D}_{k'}\rangle=\langle{\cal D}_{k}\rangle
        \langle{\cal D}_{k'}\rangle + \delta{\cal N}_{kk'} ~,
        \hspace{5mm} {\rm with}\hspace{5mm}\delta{\cal N}_{kk'}=\langle{\cal N}_{k}{\cal
        N}_{k'}\rangle - \langle{\cal N}_{k}\rangle\langle{\cal
        N}_{k'}\rangle ~.
        \label{DkDk}
        \end{equation}
The term $\delta{\cal N}_{kk'}$ can be evaluated by using the
mean-field contraction as
    \begin{equation}
        \delta{\cal N}_{kk'}\simeq
        \delta{\cal N}_{k}^{2}\delta_{kk'}~,
        \label{delta-Nkk}
    \end{equation}
with
    \begin{equation}
       \delta{\cal N}_{k}^{2} = (\delta{\cal N}_k^{+})^2 +
       (\delta{\cal N}_k^{-})^2, \hspace{5mm}
       (\delta{\cal N}_{k}^{\pm})^{2}\equiv n_k^{\pm}(1-n_k^{\pm})~,
       \label{QNF}
    \end{equation}
being the QNF for the nonzero angular momentum. The quasiparticle
occupation numbers $n_k^{\pm}$ are defined as
    \begin{equation}
       n_k^{\pm} = \langle{\cal N}_k^{\pm}\rangle ~.
       \label{nk}
    \end{equation}
From here, one can rewrite the gap equation \eqref{gapk} as a sum of
a level-independent part, $\Delta$, and a level-dependent part,
$\delta\Delta_k$, namely
    \begin{equation}
        \Delta_k = \Delta + \delta\Delta_k ~,
        \label{gapk1}
    \end{equation}
where
    \begin{equation}
        \Delta = G\sum_{k'}{u_{k'} v_{k'}\langle {\cal D}_{k'}\rangle} ~, \hspace{5mm}
        \delta\Delta_k =G\frac{\delta\mathcal{N}_k^2}{\langle {\cal D}_{k}\rangle}u_k v_k ~,
        \label{gap-final}
    \end{equation}
with
    \begin{equation}
        \langle{\cal D}_k\rangle = 1 - n_k^{+} - n_k^{-} ~.
        \label{Dk}
    \end{equation}
Within the quasiparticle mean field, the quasiparticles are
independent, therefore the quasiparticle-occupation numbers
\eqref{nk} can be approximated by the Fermi-Dirac
distribution of non-interacting fermions in the following
form
    \begin{equation}
       n_k^{\pm} = \frac{1}{{\rm exp}[\beta(E_k\mp\gamma m_k)]+1} ~.
       \label{nk12}
    \end{equation}
The equations for particle number and total angular momentum are found
by taking the average of the quasiparticle representation of
Eq. (\ref{N&M}) in the grand canonical ensemble
(\ref{average}). As the result we obtain
\begin{equation}
   N\equiv\langle\hat{N}\rangle=2\sum_{k}\bigg[v_{k}^{2}\langle{\cal
   D}_{k}\rangle +\frac{1}{2}\big(1-\langle{\cal
   D}_{k}\rangle\big)\bigg]~,
   \label{N}
\end{equation}
\begin{equation}
   M\equiv\langle\hat{M}\rangle=\sum_{k}m_{k}(n_{k}^{+}-n_{k}^{-})~.
   \label{M}
\end{equation}
We call the set of equations (\ref{gapk1}), (\ref{N}) and (\ref{M}) as
the FTBCS1 equations at finite angular momentum. By
neglecting the QNF \eqref{QNF}, as well as the screening factors
$\langle{\cal A}^{\dagger}_{k}{\cal A}_{k'}^{\dagger}\rangle$ and
$\langle{\cal A}^{\dagger}_{k}{\cal A}_{k'}\rangle$, i.e.
setting $\epsilon_k' =
\epsilon_k$ in Eq. (\ref{respe}), one recovers from Eqs.
(\ref{gapk1}), (\ref{N}) and (\ref{M})
the well-known FTBCS equations at finite
angular momentum presented in Refs. \cite{Kammuri,MorettoNPA185}.
\subsection{\label{coupling-SCQRPA}Coupling to the SCQRPA vibrations}
\subsubsection{\label{screening}SCQRPA equations and screening
factors}

The derivation of the SCQRPA equations at finite temperature and
angular momentum is carried out in the same way as that for
$T = 0$, and is formally identical to Eqs. (46), (56) and
(57) of Ref. \cite{HungDangPRC76}. The only difference is the
expressions for the screening factors
$\langle\mathcal{A}_k^+\mathcal{A}_{k'}^+\rangle$ and
$\langle\mathcal{A}_k^+\mathcal{A}_{k'}\rangle$ at the right-hand
side of Eq. \eqref{respe}, which are now the functions
of not only the SCQRPA amplitudes,
but also of the expectation values $\langle
Q_{\mu}^+ Q_{\mu '}\rangle$ and $\langle Q_{\mu}^+ Q_{\mu
'}^+\rangle$ of the SCQRPA operators.
As the details of the derivation are given in
Ref. \cite{DangHungPRC2008}, only final expressions are quoted
below. The screening factors are given as
    \begin{equation}
        x_{kk'}\equiv\langle\bar{\cal A}_{k}^{\dagger}
        \bar{\cal A}_{k'}\rangle =\sum_{\mu}{\cal Y}_{k}^{\mu}{\cal Y}_{k'}^{\mu}
        +\sum_{\mu\mu'} \bigg(U_{kk'}^{\mu\mu'}\langle{\cal
        Q}^{\dagger}_{\mu} {\cal Q}_{\mu'}\rangle +
        Z_{kk'}^{\mu\mu'}\langle{\cal Q}^{\dagger}_{\mu} {\cal
        Q}_{\mu'}^{\dagger}\rangle\bigg)~, \label{xkk}
    \end{equation}
    \begin{equation}
        y_{kk'}\equiv\langle\bar{\cal A}_{k}^{\dagger}
        \bar{\cal A}_{k'}^{\dagger}\rangle=\sum_{\mu}{\cal
        Y}_{k}^{\mu}{\cal X}_{k'}^{\mu} +\sum_{\mu\mu'}
        \bigg(U_{kk'}^{\mu\mu'}\langle{\cal Q}^{\dagger}_{\mu} {\cal
        Q}_{\mu'}^{\dagger}\rangle + Z_{kk'}^{\mu\mu'}\langle{\cal
        Q}^{\dagger}_{\mu} {\cal Q}_{\mu'}\rangle\bigg)~, \label{ykk}
    \end{equation}
where
    \begin{equation}
        \bar{\cal A}_{k}^{\dagger}=
           \frac{{\cal A}_{k}^{\dagger}}{\sqrt{\langle{\cal
           D}_{k}\rangle}}~,\hspace{5mm} \bar{\cal A}_{k}=
           [\bar{\cal A}_{k}^{\dagger}]^{\dagger}~,\hspace{5mm}
{U}_{kk'}^{\mu\mu'}={\cal X}_{k}^{\mu}{\cal X}_{k'}^{\mu'}
        +{\cal Y}_{k'}^{\mu}{\cal
        Y}_{k}^{\mu'}~,\hspace{5mm}
        {Z}_{kk'}^{\mu\mu'}={\cal X}_{k}^{\mu}{\cal Y}_{k'}^{\mu'}
        +{\cal Y}_{k}^{\mu'}{\cal X}_{k'}^{\mu}~,
        \label{UW}
    \end{equation}
with ${\cal X}_{k}^{\mu}$ and ${\cal Y}_{k}^{\mu}$ being the
amplitudes of the SCQRPA operators~\footnote{Although at
finite angular momentum, the expectation value
$\langle[{\cal B}_{k}, {\cal B}_{k'}^{\dagger}]\rangle=
\delta_{kk'}(n_{k}^{-}-n_{k}^{+})$, where ${\cal
B}^{\dagger}_{k}\equiv\alpha_{k}^{\dagger}\alpha_{-k}$,
becomes finite as $n_{k}^{-}\neq
n_{k}^{+}$, the scattering operators
${\cal B}_{k}$ and ${\cal B}_{k}^{\dagger}$ do
not contribute to the QRPA because they
commute with operators ${\cal A}_{k}^{\dagger}$, ${\cal A}_{k}$, and
${\cal N}_{k}$.}
    \begin{equation}
        {\cal Q}_{\mu}^{\dagger}=\sum_{k}
        ({\cal X}_{k}^{\mu}
        \bar{\cal A}_{k}^{\dagger}-{\cal Y}_{k}^{\mu}
        \bar{\cal A}_{k})~,\hspace{5mm} {\cal Q}_{\mu}=[{\cal
        Q}_{\mu}^{\dagger}]^{\dagger}~.
        \label{Qmu}
    \end{equation}
The expectation values of $\langle {\cal Q}_{\mu}^{\dagger}{\cal Q}_{\mu '}\rangle$
and $\langle{\cal Q}_{\mu}^{\dagger}{\cal Q}_{\mu '}^{\dagger}\rangle$
are found as
    \begin{equation}
        \langle{\cal Q}_{\mu}^{\dagger}{\cal
        Q}_{\mu'}\rangle =\sum_{k}{\cal Y}_{k}^{\mu}{\cal Y}_{k}^{\mu'}+
        \sum_{kk'}(U_{kk'}^{\mu\mu'}x_{kk'}-W_{kk'}^{\mu\mu'}y_{kk'})~,
        \label{Q+Q}
    \end{equation}
    \begin{equation}
        \langle{\cal Q}_{\mu}^{\dagger}{\cal
        Q}_{\mu'}^{\dagger}\rangle =-\sum_{k}{\cal Y}_{k}^{\mu}{\cal
        X}_{k}^{\mu'}+
        \sum_{kk'}(U_{kk'}^{\mu\mu'}y_{kk'}-W_{kk'}^{\mu\mu'}x_{kk'})~,
        \label{Q+Q+}
    \end{equation}
where
    \begin{equation}
        W^{\mu\mu'}_{kk'}={\cal X}_{k}^{\mu}{\cal Y}_{k'}^{\mu'}
        +{\cal Y}_{k'}^{\mu}
        {\cal X}_{k}^{\mu'}~.
        \label{W}
    \end{equation}
From Eqs. \eqref{xkk}, \eqref{ykk}, \eqref{Q+Q} and \eqref{Q+Q+}, the set of exact equations for
the screening factors is obtained in the form
    \[
        \sum_{k_{1}k'_{1}}\bigg[\delta_{kk_{1}}\delta_{k'k_{1}'}
        -\sum_{\mu\mu'}\big(U_{kk'}^{\mu\mu'}U_{k_{1}k_{1}'}^{\mu\mu'}
        -Z_{kk'}^{\mu\mu'}W_{k_{1}k_{1}'}^{\mu\mu'}\big)\bigg]x_{k_{1}k_{1}'}
        +\sum_{k_{1}k_{1}'\mu\mu'}\big(U_{kk'}^{\mu\mu'}W_{k_{1}k_{1}'}^{\mu\mu'}
        -Z_{kk'}^{\mu\mu'}U_{k_{1}k_{1}'}^{\mu\mu'}\big)y_{k_{1}k_{1}'}
    \]
    \begin{equation}
        =\sum_{\mu}{\cal Y}_{k}^{\mu}{\cal Y}_{k'}^{\mu}
        +\sum_{k''\mu\mu'}{\cal Y}_{k''}^{\mu}
        \big(U_{kk'}^{\mu\mu'}{\cal Y}_{k''}^{\mu'}
        -Z_{kk'}^{\mu\mu'}{\cal X}_{k''}^{\mu'}\big)~,
        \label{x1}
    \end{equation}
    \[
        \sum_{k_{1}k_{1}'\mu\mu'}\big(U_{kk'}^{\mu\mu'}W_{k_{1}k_{1}'}^{\mu\mu'}
        -Z_{kk'}^{\mu\mu'}U_{k_{1}k_{1}'}^{\mu\mu'}\big)x_{k_{1}k_{1}'}
        +\sum_{k_{1}k'_{1}}\bigg[\delta_{kk_{1}}\delta_{k'k_{1}'}
        -\sum_{\mu\mu'}\big(U_{kk'}^{\mu\mu'}U_{k_{1}k_{1}'}^{\mu\mu'}
        -Z_{kk'}^{\mu\mu'}W_{k_{1}k_{1}'}^{\mu\mu'}\big)\bigg]y_{k_{1}k_{1}'}
    \]
    \begin{equation}
        =\sum_{\mu}{\cal Y}_{k}^{\mu}{\cal X}_{k'}^{\mu}
        +\sum_{k''\mu\mu'}{\cal Y}_{k''}^{\mu}
        \big(Z_{kk'}^{\mu\mu'}{\cal Y}_{k''}^{\mu'}
        -U_{kk'}^{\mu\mu'}{\cal X}_{k''}^{\mu'}\big)~.
        \label{y1}
    \end{equation}

\subsubsection{\label{occupation}Quasiparticle occupation numbers}

The quasiparticle occupation numbers
\eqref{nk} are calculated by coupling to the SCQRPA phonons making use
of the method of double-time Green's functions~\cite{Bogoliubov,Zubarev}.
By representing the Hamiltonian \eqref{QHa} in the effective
form as
    \begin{equation}
        H_{eff}=\sum_{k}b_k^{+}{\cal N}_k^{+} + \sum_{-k}b_k^{-}{\cal N}_k^{-}
        +\sum_{k'}q_{kk'}{\cal N}_k{\cal N}_{k'}
        +\sum_{\mu}\omega_{\mu}{\cal Q}_{\mu}^{\dagger}{\cal Q}_{\mu}+\sum_{k\mu}V_{k}^{\mu} {\cal
        N}_{k}({\cal Q}_{\mu}^{\dagger}+{\cal Q}_{\mu})~.
        \label{Heff}
    \end{equation}
with $\omega_\mu$ denoting the phonon energies (eigenvalues of the SCQRPA
equations) and the vertex $V_k^{\mu}$ given as
    \begin{equation}
        V_{k}^{\mu}=\sum_{k'}g_{k}(k')\sqrt{\langle{\cal D}_{k'}\rangle}
        ({\cal X}_{k'}^{\mu}+{\cal Y}_{k'}^{\mu})~,
        \label{Vertex}
    \end{equation}
we introduce the following double-time Green's functions for the
quasiparticle propagations
    \begin{equation}
        G_{\pm k}(t-t')=\langle\langle\alpha_{\pm
        k}(t);\alpha^{\dagger}_{\pm k}(t')\rangle\rangle~,
        \label{Gk}
    \end{equation}
    as well as those corresponding to quasiparticle$\otimes$phonon couplings
    \begin{equation}
        {\Gamma}_{{\pm k}\mu}^{--}(t-t')=\langle\langle\alpha_{\pm k}(t){\cal
        Q}_{\mu}(t);\alpha^{\dagger}_{\pm k}(t')\rangle\rangle~,\hspace{5mm}
        {\Gamma}_{{\pm k}\mu}^{-+}(t-t')=\langle\langle\alpha_{\pm k}(t) {\cal
        Q}_{\mu}^{\dagger}(t);\alpha^{\dagger}_{\pm k}(t')\rangle\rangle~.
        \label{Gamma}
    \end{equation}
Following the same procedure in Ref.
\cite{DangHungPRC2008}, we obtain the final equations for the
    quasiparticle Green's functions $G_{\pm k}(E)$ in the
    following form
    \begin{equation}
        G_{\pm k}(E)=\frac{1}{2\pi}\frac{1}{E-E_{k}^{\pm}-M_{k}^{\pm}(E)}~,
        \label{GkE}
    \end{equation}
where
    \begin{equation}
        E_k^{\pm} = b_k^{\pm} + q_{kk}~,
        \label{Ek+-}
        \end{equation}
        \begin{equation}
        M_{k}^{\pm}(E=\omega\pm i\varepsilon)=M_{k}^{\pm}(\omega)\mp i\gamma_{k}^{\pm}(\omega)~,
        \label{Mk}
    \end{equation}
    \begin{equation}
        M_{k}^{\pm}(\omega)=\sum_{\mu}(V_{k}^{\mu})^{2}
        \bigg[\frac{(1-n_{k}^{\pm}+\nu_{\mu})(\omega-E_{k}^{\pm}-\omega_{\mu})}
        {(\omega-E_{k}^{\pm}-\omega_{\mu})^{2}+\varepsilon^{2}}+
        \frac{(n_{k}^{\pm}+\nu_{\mu})(\omega-E_{k}^{\pm}+\omega_{\mu})}
        {(\omega-E_{k}^{\pm}+\omega_{\mu})^{2}+\varepsilon^{2}}\bigg]~,
        \label{Momega}
    \end{equation}
    \begin{equation}
        \gamma_{k}^{\pm}(\omega)=\varepsilon
        \sum_{\mu}(V_{k}^{\mu})^{2}
        \bigg[\frac{1-n_{k}^{\pm}+\nu_{\mu}}
        {(\omega-E_{k}^{\pm}-\omega_{\mu})^{2}+\varepsilon^{2}}+
        \frac{n_{k}^{\pm}+\nu_{\mu}}
        {(\omega-E_{k}^{\pm}+\omega_{\mu})^{2}+\varepsilon^{2}}\bigg]~.
        \label{gamma}
    \end{equation}
In Eqs. (\ref{Mk}) -- (\ref{gamma}),
the imaginary parts $\gamma^{\pm}_{k}(\omega)$ ($\omega$ real) of the
analytic continuation of $M_{k}^{\pm}(E)$ into the complex energy describe the
damping of quasiparticle excitations due to coupling to SCQRPA
vibrations, $\nu_\mu = \langle{\cal
Q}_{\mu}^{+}{\cal Q}_{\mu}\rangle$ is the phonon occupation number,
and $\varepsilon$ is a sufficient small parameter.
These results allow to find the spectral intensities
$J_{k}^{\pm}(\omega)$ from the relations $J_{k}^{\pm}(\omega)=i[G_{\pm k}(\omega+i\varepsilon)-G_{\pm
k}(\omega-i\varepsilon)]/[\exp(\beta\omega)+1]$ in the form
\begin{equation}
J_{k}^{\pm}(\omega)=\frac{1}{\pi}
        \frac{\gamma_{k}^{\pm}(\omega)(e^{\beta\omega}+1)^{-1}}
        {[\omega-E_{k}^{\pm}-M_{k}^{\pm}(\omega)]^{2}+[\gamma_{k}^{\pm}(\omega)]^2}~,
        \label{Jk+-}
        \end{equation}
and, finally, the quasiparticle
occupation numbers \eqref{nk} as
    \begin{equation}
        n_{k}^{\pm}=\int_{-\infty}^{\infty}J_{k}^{\pm}(\omega)d\omega~.
        \label{nkcoupling}
    \end{equation}
In the limit of quasiparticle damping
$\gamma_k^{\pm}(\omega)\rightarrow 0$, $n_k^{\pm}$ can be
approximated with the Fermi-Dirac distribution
    \begin{equation}
        n_{k}^{\pm}\simeq\frac{1}{{\rm exp}(\beta \widetilde{E}_{k}^{\pm})+1}~,
        \label{nklimit}
    \end{equation}
where $\widetilde{E}_{k}^{\pm}$ are the solutions of the equations for the
poles
of the quasiparticle Green's functions $G_{\pm k}(\omega)$ (\ref{GkE}), namely
    \begin{equation}
        \widetilde{E}_{k}^{\pm}-E_{k}^{\pm}-M_{k}^{\pm}(\widetilde{E}_{k}^{\pm}) = 0~.
    \end{equation}

The particle-number violation inherent in the BCS-based theories still
causes some quantal fluctuation of particle number starting from $T=$ 0.
This defect can be removed by
carrying out a proper particle-number projection (PNP).
Among different methods of PNP, the Lipkin-Nogami (LN) prescription
(LN-PNP)~\cite{LN} is widely used because of its
simplicity. This method has been implemented into the
FTBCS1 and FTBCS1+SCQRPA in Ref.~\cite{DangHungPRC2008}, and the ensuing
approaches are called the FTLN1 and FTLN1+SCQRPA, respectively. Their
extension to $M\neq$ 0 is straightforward.
It is easy to see that, in the nonrotating limit ($\gamma=$ 0), one has
$b_{k}^{+}=b_{k}^{-}=b_{k}$ from Eq. (\ref{bk}),
$n_{k}^{+}=n_{k}^{-}$ from Eqs. (\ref{nkcoupling}), and all
above-derived formalism reduces to that presented in Ref.
\cite{DangHungPRC2008}.
\section{\label{sec3}NUMERICAL RESULTS}
\subsection{Ingredients of numerical calculations}
The numerical calculations are carried out within a schematic model
as well as several nuclei with realistic single-particle spectra. For
the schematic model, we use the one with $N$ particles distributed
over $\Omega=N$ doubly-folded equidistant
levels.
These levels interact
via an attractive pairing force with the constant parameter
$G$. When the interaction is switched off, all the lowest
$\Omega/2$ levels are filled up with $N$ particles so that each
of them is occupied by two
particles with the spin projections equal to $\pm m_{k}$
($k$=1,\ldots, $\Omega$, and $m_k =$ 1/2, 3/2, ... , $\Omega-1/2$).
The single-particle energies $\epsilon_{k}$
are measured from the middle of the spectrum as $\epsilon_{k}=\epsilon[k-(\Omega+1)/2]$ so
that the energies of the lower 5 levels are negative, whereas those of the
upper 5 levels are positive.
The results
obtained for $N=$ 10, $\epsilon=$ 1 MeV, and $G=$ 0.5 MeV are analyzed in the present
paper.

As for the realistic nuclei, we carry out the calculations for neutrons
in $^{20}$O and $^{44}$Ca, whereas
the contribution of proton and neutron components to nuclear
pairing is studied for the well-deformed $^{22}$Ne nucleus,
where a backbending of moment of inertia as a function of the square
of angular velocity was detected~\cite{SzantoPRL42}. The calculations use
the single-particle energies generated at $T=$ 0 within deformed
Woods-Saxon potentials.  For the slightly axially
deformed $^{20}$O, the multipole deformation parameters $\beta_2$, $\beta_3$, $\beta_4$,
$\beta_5$, and $\beta_6$ are chosen to be equal to
0.03, 0.0, -0.108, 0.0, and -0.003, respectively.  For $^{22}$Ne, the
axial deformation is rather strong with these parameters taking the
values equal to 0.326, 0.0, 0.225, 0.0, and 0.011, respectively.
For the spherical $^{44}$Ca, all the deformation parameters $\beta_{i}$
are set to
be equal to zero.
Other parameters of Woods-Saxon potentials are taken from Table 1 of Ref.
\cite{CwiokCPC46}. The neutron
single-particle spectrum for $^{20}$O includes all levels
up to the shell closure with $N=$ 20 (between around -25.84 MeV and 0.49
MeV), from which two orbitals, $1d_{3/2}$ and
$1d_{1/2}$, are unbound. These unbound states
have been shown to have a large contribution to
pairing correlations in $^{20-22}$O isotopes~\cite{BetanNPA771}.
The neutron single-particle spectrum for $^{44}$Ca include all
bound states between around -35.6 MeV and -1.05 MeV,
up to the $2p_{1/2}$ orbital of the closed shell with
$N=$ 50. The single-particle spectra for $^{22}$Ne consist of all
11 proton bound states between
-30.23 $\leq\epsilon_{p}\leq$ -0.156 MeV, and 12 neutron ones between
-29.834 $\leq\epsilon_{n}\leq$ -0.742 MeV.
The values of pairing interaction parameter $G$
are chosen so that the pairing gaps $\Delta(T=0, M=0)$ obtained at
zero temperature and zero angular momentum
match the experimental values extracted from the odd-even mass
differences for these nuclei, namely, $\Delta(0, 0)\simeq$
4 MeV for protons in $^{22}$Ne, and 3, 2, and 3 MeV
for neutrons in $^{20}$O, $^{44}$Ca,
$^{22}$Ne, respectively.

The numerical calculations are carried out within the FTBCS and
FTBCS1 for the level-weighted pairing gap
$\bar\Delta=\sum_k{\Delta_k}/\Omega$ as functions of temperature $T$,
angular momentum $M$, and angular velocity
$\gamma$. The effect caused by coupling to
SCQRPA vibrations is analyzed by studying the
total energy ${\cal E} = \langle H \rangle$ and heat
capacity $C =\partial {\cal E} / \partial T$, whereas the backbending is
studied by considering the momentum of inertia as a function of
$\gamma^{2}$ as $T$ varies.
\subsection{\label{model}Results within the doubly-folded multilevel
equidistant model}
    \begin{figure}
        \includegraphics[width=10cm]{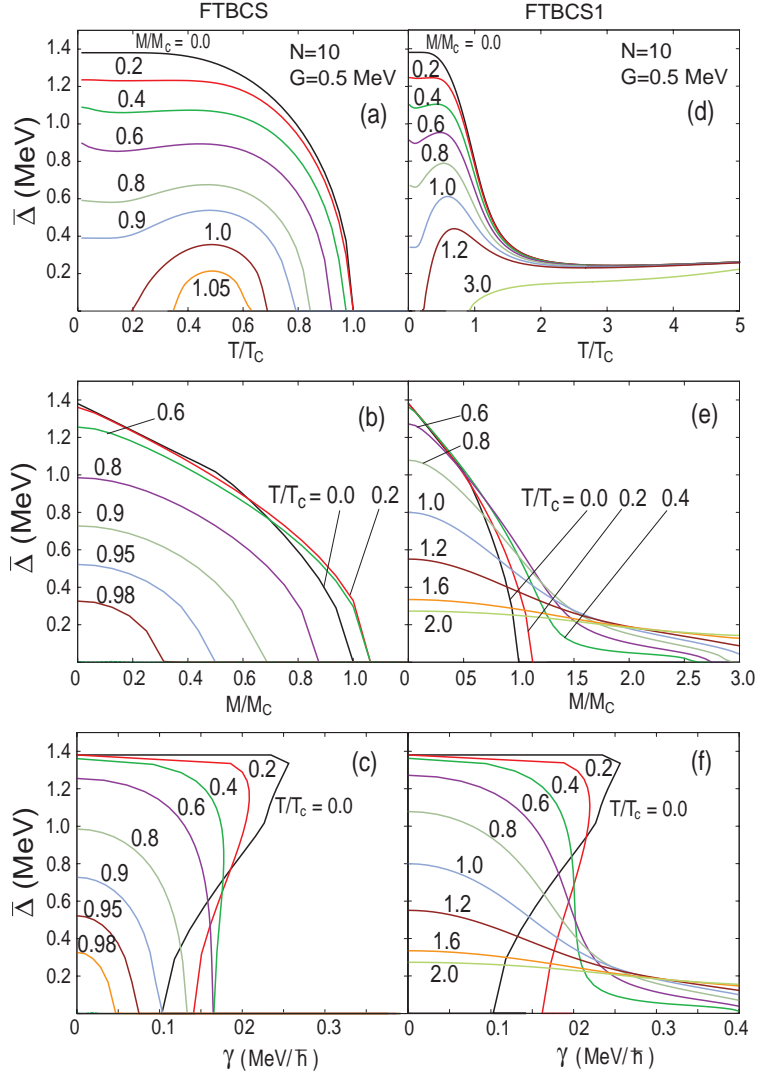}
        \caption{(Color online)  Level-weighted pairing gaps
        $\bar{\Delta}$ as
        functions of $T$ at various $M$ [(a), (d)], and as functions of
        $M$ [(b), (e)] and $\gamma$ [(c), (f)] at several $T$ for
        $N$ = 10, $G$ = 0.5 MeV obtained within the FTBCS (left) and
        FTBCS1 (right).
        \label{N10}}
    \end{figure}
Shown in Figs. \ref{N10} (a) and \ref{N10} (d)
are the level-weighted pairing gaps $\bar{\Delta}$ as
functions of $T$ at various $M$, whereas the dependence of $\bar{\Delta}$
on $M$  at several $T$ is displayed in Figs. \ref{N10} (b)
and \ref{N10}(e). Finally,
Figs. \ref{N10} (c) and \ref{N10} (f) show the gaps $\bar{\Delta}$ as
functions of the angular velocity $\gamma$ at various $T$.
All the results are obtained for the system with
$N$ = 10 and $G$ = 0.5 MeV, from which the left panels are the
predictions by the FTBCS theory, whereas the right panels are those by
the FTBCS1 one. It is clearly seen from Figs. \ref{N10} (a) and
\ref{N10} (b) that the FTBCS gap
decreases with increasing $T$ ($M$) at $M=$ 0 ($T=$ 0) up to a certain
critical value $T_{\rm c}=$ 0.77 MeV ($M_{\rm c}=$ 8$\hbar$),
where the FTBCS gap
collapses. The collapse of the pairing gap at $M=M_{\rm c}$ (at $T=$
0) was proposed by Mottelson and Valatin as being caused by the
Coriolis force, which breaks the Cooper pairs~\cite{MV}.
This feature remains with the FTBCS gap as a function of
$M$, when $T\neq$ 0, but with decreasing $M_{\rm c}(T)<M_{\rm c}$ as
$T$ increases beyond 0.6$T_{\rm c}$.
As for the behavior of the FTBCS gap as a function of
$T$,  one notices that, at $M$ slightly larger than $M_{\rm c}$,
the so-called thermally assisted pairing correlation takes place, in which
the pairing gap is zero at $T \leq T_1$, increases at
$T>T_1$ to reach a maximum, then decreases again to vanish
at $T\geq T_2$ [See. Fig. \ref{N10} (a) for $M/M_{\rm c}\geq$ 1].
This interesting phenomenon was predicted  and explained,
for the first time, by Moretto in Refs. \cite{MorettoPLB35,MorettoNPA185}
by applying the FTBCS to the uniform model. At
$M/M_{\rm c}>$ 1.1, no FTBCS pairing gap remains.

Different from the FTBCS predictions,
the results obtained within the FTBCS1 include the effect caused by
the QNF. As one can see in Fig. \ref{N10} (d), while, in the
region of low temperature $T < T_{\rm c}$, the FTBCS1 and FTBCS gaps for
different $M$ are rather similar, they are qualitatively different  at
$T > T_{\rm c}$. Here, the QNF, which is
rather strong at high $T$, causes a monotonous decrease of
the FTBCS1 gap $\bar{\Delta}$ as $T$
increases. This FTBCS1 gap never collapses even at very high $T$.
Instead all the values of the FTBCS1 gap obtained at various $M$
seem to saturate at a value of around 2.25 MeV at $T>$ 5 MeV.
This feature shows that, the effect of angular momentum
on reducing the pairing correlation is significant only at low $T$.
In the high temperature
region, the QNF leads to a persistence of the pairing correlation
in a rotating system. Compared to the FTBCS theory, when the QNF is
neglected, the effect of thermally assisted
pairing correlation also takes place at $M/M_{\rm c} > 1.1$. However,
the FTBCS1 gap is now zero at $T \leq T_1$, reappears at $T
= T_1$, and remains finite at $T > T_1$.
This result is found in qualitative agreement with
those obtained in the calculations of the
canonical gap of in Ref. \cite{FrauendorfPRB68}, where the
reappearance of the pairing gap at finite $T$ and $\gamma$ is related
to the strong fluctuations of order parameter in the canonical ensemble
of small systems such as metal clusters and nuclei. In the present
paper, we point out the QNF as the microscopic origin of this effect.
Comparisons between the FTBCS1 gaps and the canonical ones obtained
at ($T\neq$ 0, $M=$ 0) and ($T=$ 0, $M$ and/or $\gamma \neq$ 0) are discussed later,
in Sec. \ref{correction}.

The QNF has a similar effect on the behavior of the pairing gap
$\bar{\Delta}$ as a function of angular momentum.
As low $T$, when the QNF is still negligible,
the FTBCS and FTBCS1 gaps as functions of $M$ are similar. They both
decreases as $M$ increases, and collapse at $M=M_{\rm c}$
and at $M$ slightly higher than $M_{\rm c}$
for  0 $< T/T_{\rm c} \leq$ 0.2, contrary to the trend within the
FTBCS theory, where $M_{\rm c}(T)$ decreases
as $T/T_{\rm c}$ increases above 0.6 discussed above [Compare
Figs. \ref{N10} (b) and \ref{N10} (e)].
At $T/T_{\rm c} =$ 0.8, e.g., the collapsing points of the FTBCS and
FTBCS1 gaps are $M/M_{\rm c}
\simeq$ 0.85, and 2.9, respectively.

The FTBCS and FTBCS1 pairing
gaps are displayed in Figs. \ref{N10} (c) and \ref{N10} (f)
as functions of angular velocity $\gamma$  at various $T$.
For $T/T_{\rm c} \leq$ 0.2, the
pairing gap undergoes a backbending, which will be
discussed in the Sec. \ref{backbend1}. At $T/T_{\rm c} >$ 0.2
no backbending is seen for the pairing gaps.
This result agrees with those obtained in calculations within the
finite-temperature Hartree-Fock-Bogoliubov cranking (FTHFBC) theory,
which is applied to the two-level model in Ref. \cite{GoodmanNPA352}.
Within the FTBCS1, the
pairing gaps at large $M$ become enhanced with
$T$, in agreement with the results obtained within an
exactly solvable model for a single $f_{7/2}$ shell
in Ref. \cite{SheikhPRC72}.
\subsection{\label{realistic}Results for realistic nuclei}
    \begin{figure}
        \includegraphics[width=10cm]{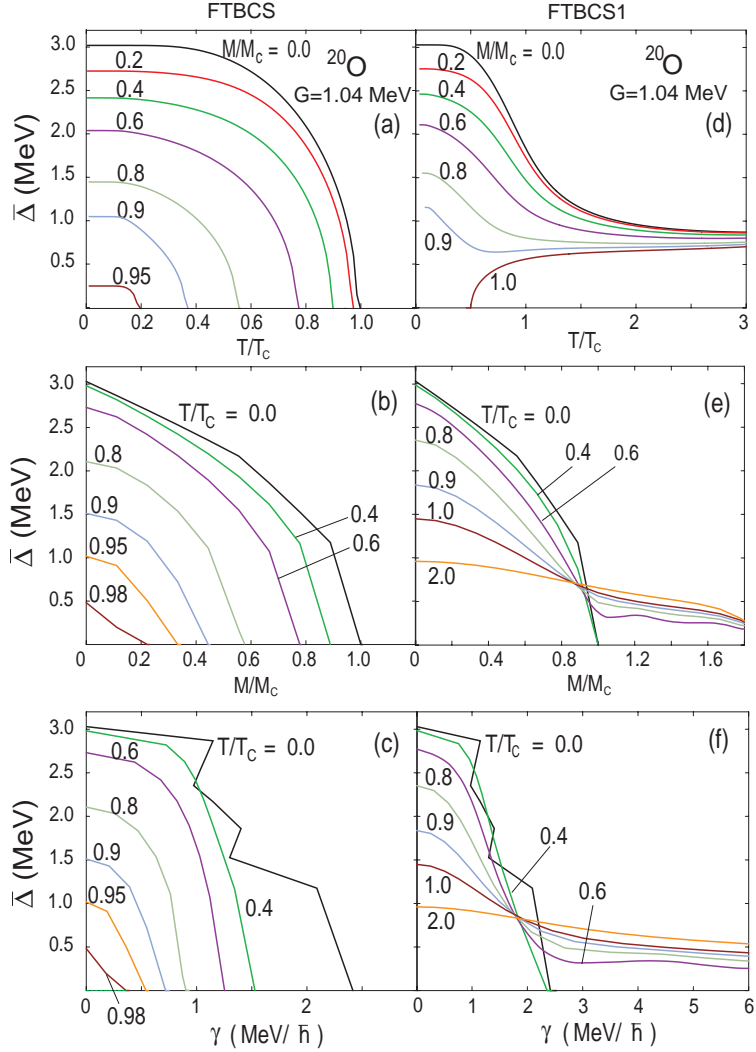}
        \caption{(Color online)  Same as
        Fig. \ref{N10} but for neutrons in $^{20}$O
        using $G =$ 1.04 MeV.
        \label{O20}}
    \end{figure}
Shown in Fig. \ref{O20} are the level-weighted pairing gaps as functions of
$T, M$ and $\gamma$ obtained within the FTBCS and FTBCS1 theories for
neutrons in $^{20}$O. The values of $T_{\rm c}$ (at $M=$ 0) and $M_{\rm c}$
(at $T=$ 0) are
found equal to 1.57 MeV
and 4$\hbar$, respectively. Compared to the case with schematic model
discussed in the previous section, the
difference is that no thermally
assisted pairing correlation appears within the FTBCS for $^{20}$O.
All the FTBCS gaps behave similarly as functions of $T$ with
increasing $M$.  At a given value of $M$,
they decrease with increasing both $T$, and collapse at
some values $T_{\rm c}(M)<T_{\rm c}$. A similar behavior is seen for the gaps
as functions of $M$ at a given value of $T$. Here the critical value
$M_{\rm c}(T)$ for the angular momentum, at which the gap collapses is found
decreasing with increasing $T$ so that $M_{\rm c}(T)<M_{\rm c}$ [See Figs. \ref{O20}
(a) and \ref{O20} (b)].
Meanwhile, the temperature dependence of the FTBCS1 gap in Fig.
\ref{O20} (d) shows a clear
manifestation of the thermally assisted pairing gap. As $M$ increases
up to $M/M_{\rm c}\simeq$ 0.8, the gap decreases monotonously with
increasing $T$ up to $T\simeq$ 1.5$T_{\rm c}$, higher than which the
gap seems to be rather stable against the variation of $T$.
At $M/M_{\rm c}\geq$ 0.9, the reentrance of thermal pairing starts to
show up as the enhancement of the tail at $T>T_{\rm c}$. When
$M/M_{\rm c}$ becomes equal to or larger than 1, the gap completely
vanishes at low $T$, but reappears starting from a certain value of
$T$, above which the gap increases with $T$ and reach a saturation at
high $T$. At $T\geq$ 3 MeV, all the gaps obtained at different values
of $M$ seem to coalesce to limiting value around 0.7 -- 0.8 MeV.
At a given value of $T$ in the region $T/T_{\rm c}\leq$ 0.7,
as shown in Fig. \ref{O20} (e), the
pairing gaps decrease steeply with increasing $T$ and all collapse at
the same value $M_{\rm c}$. This difference compared to the FTBCS
theory comes from the presence of the QNF. At $T/T_{\rm c}\geq$
0.8, the QNF becomes stronger, which pushes up the
collapsing point to $M_{\rm c}(T)= 2M_{\rm c}$. One can also sees some
oscillation occurring in the region between 0.8 $\leq M/M_{\rm c}\leq$
1.4 because of the shell structure. The collapsing point might be
shifted even further to higher $M$ with increasing $T$, but at too
high $T$ the temperature dependence of single-particle energies
becomes significant so that the use of the spectrum obtained at $T=$ 0
is no longer valid~\cite{BrackBonche}.

The pairing gaps as
functions of angular velocity $\gamma$ obtained at various $T$ within the FTBCS and
FTBCS1 theories are plotted in Figs.
\ref{O20} (c) and \ref{O20} (f), respectively.
As $E_{k}$, $\gamma$ and $m_{k}$ are positive, at $T = 0$, the quasiparticle occupation
number $n_k^-$ is always zero, whereas $n_k^+$ is a step function of
$E_{k}-\gamma m_{k}$, which is zero if $E_k > \gamma m_k$ and
1 if $E_k \leq \gamma m_k$.
As the result, the FTBCS and FTBCS1 gaps decrease with increasing
$\gamma$ in a stepwise manner up to a critical value $\gamma_{\rm c}$, where they
vanish. At $T\neq$ 0, the Fermi-Dirac distribution replaces the step
function, which washes out the
stepwise manner in the behavior of the gaps as functions of the
$\gamma$. Here again, once can see that, at $T/T_{\rm c}>$ 0.8, the QNF is so
strong that the collapse of the FTBCS1 gap is  completely smoothed out
[Fig. \ref{O20} (f)].

    \begin{figure}
        \includegraphics[width=10cm]{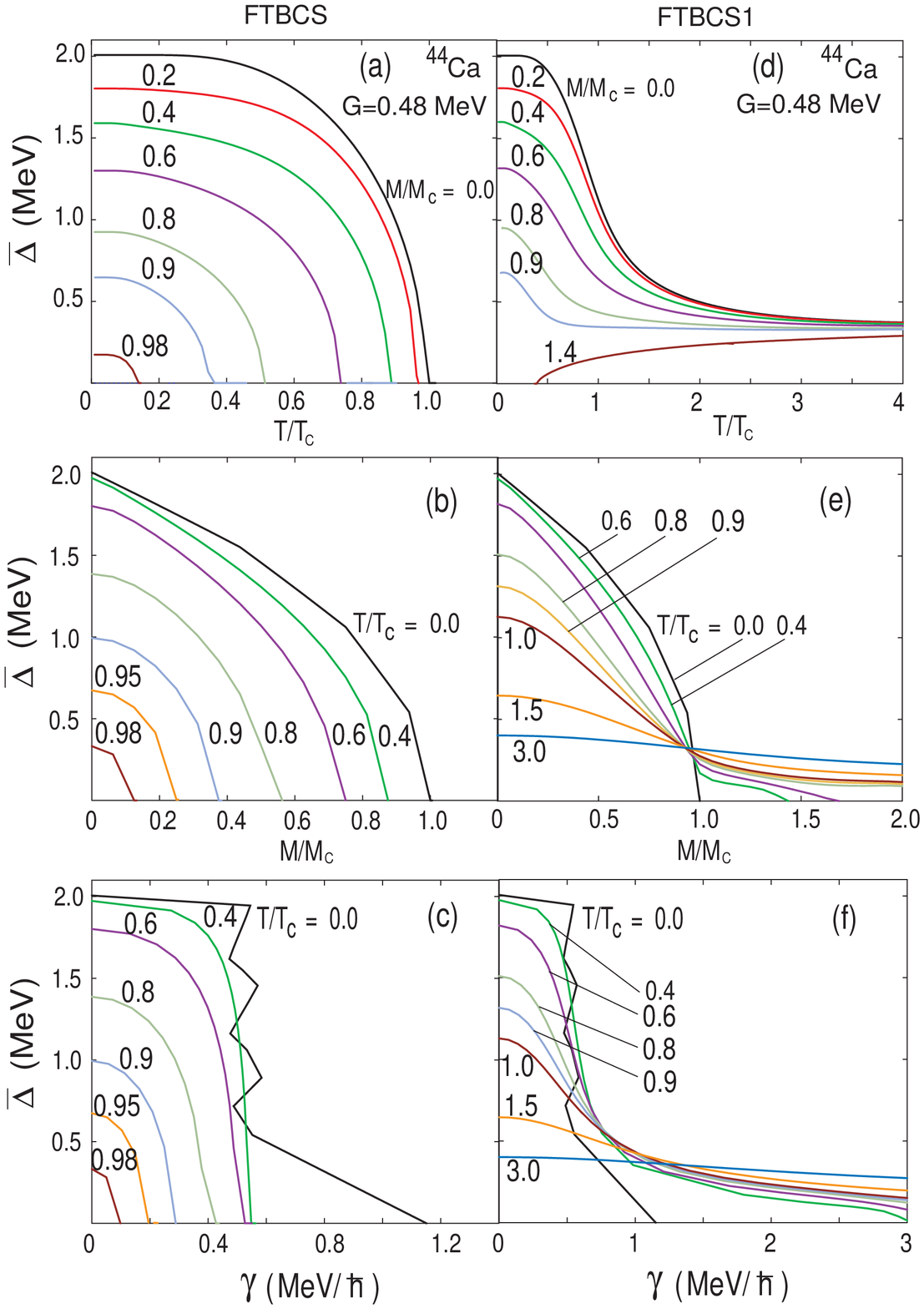}
        \caption{(Color online)  Same as Fig. \ref{N10}
        but for neutrons in $^{44}$Ca
        using $G =$ 0.48 MeV.
        \label{Ca44}}
    \end{figure}
The level-weighted pairing gaps $\bar{\Delta}$ for neutrons in $^{44}$Ca
shown in Fig. \ref{Ca44} have a similar behavior as as
functions of $T$, $M$ and $\gamma$ with the values of $T_{\rm c}$ and
$M_{\rm c}$ are found to be 1.07 MeV
and 8 $\hbar$, respectively. The
thermally assisted pairing  gap appears at $M/M_{\rm c} >$ 1.0 but
the high-$T$ tail is much depleted due to a weaker
QNF in a heavier system compared to that in $^{20}$O.

    \begin{figure}
        \includegraphics[width=10cm]{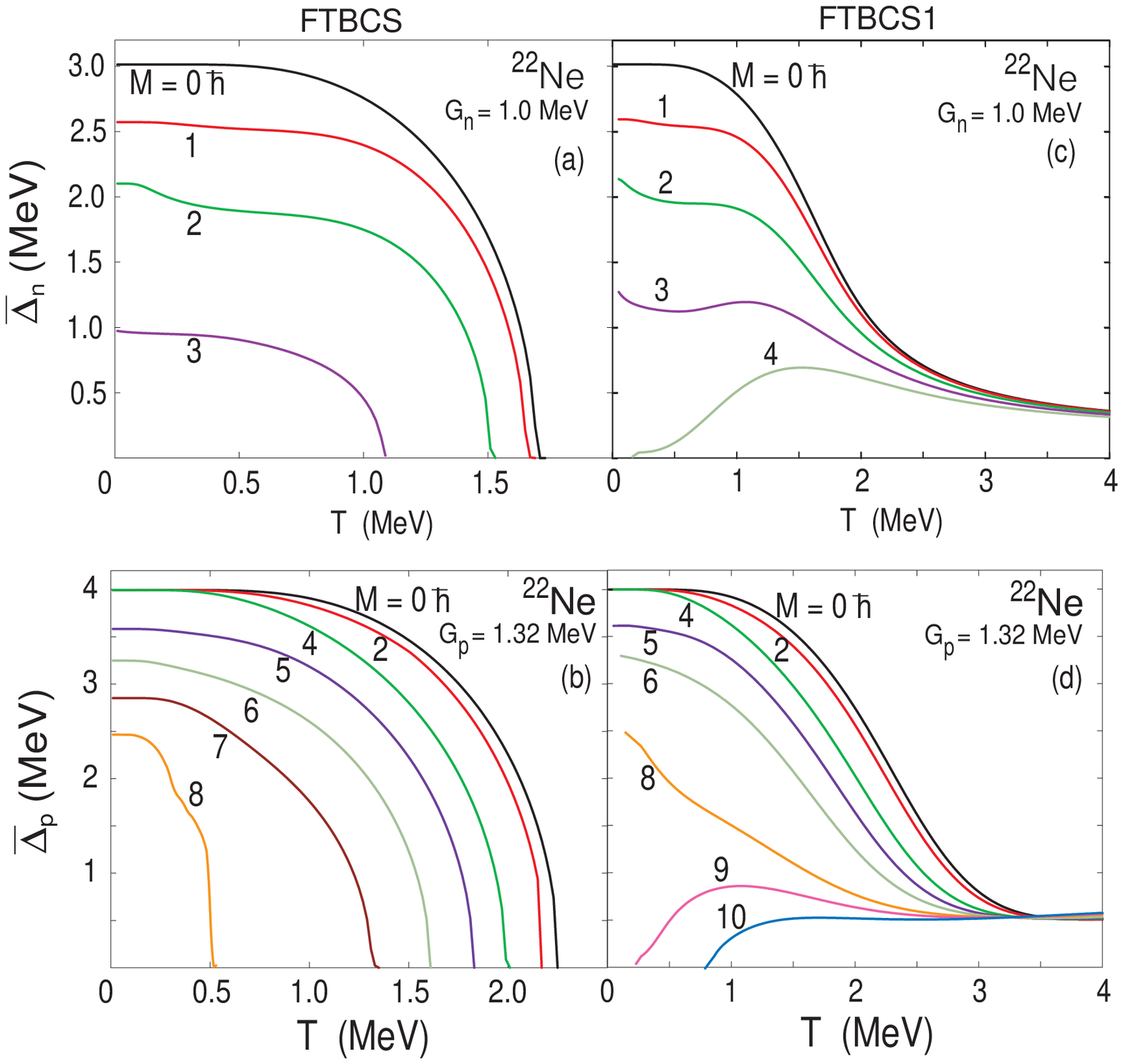}
        \caption{(Color online) Level-weighted pairing gaps
        as functions
        of $T$ at various $M$ obtained within the FTBCS (left) and
        FTBCS1 (right) for neutrons [(a), (c)],
        and protons [(b), (d)] in $^{22}$Ne
        using $G_n =$ 1.0 MeV and $G_p =$ 1.32 MeV.
        \label{Ne22}}
    \end{figure}
The well deformed nucleus $^{22}$Ne has both neutron and proton open
shells, therefore the gap and two number equations for
protons ($p$) and neutrons ($n$) are simultaneously
solved together with one equation for the total
angular momentum $M = M_{p} + M_{n}$ to
obtain the pairing gaps $\Delta_{p}$ and  chemical
potential $\lambda_{p}$ for protons, the corresponding
quantities, $\Delta_{n}$ and $\lambda_{n}$, for neutrons, as well
as the angular velocity $\gamma$ of the entire nucleus~\cite{MorettoNPA216}.
The level-weighted pairing gaps as functions
of $T$ at several $M$ obtained for neutrons and protons in $^{22}$Ne are
shown in Fig. \ref{Ne22}. The FTBCS neutron gaps
become depleted with increasing $M$, and completely disappears at $M > 3\hbar$.
As a function of $T$, the FTBCS neutron gaps
decrease as $T$ increases and collapse at $T_{\rm c}(M)$,
which decreases from $T_{\rm c}(M=0)\simeq$ 1.7 MeV to $T_{\rm
c}(M=3\hbar)\simeq$ 1.1 MeV. The FTBCS1 gaps obtained at $M<$ 3
$\hbar$ never collapse, but gradually decrease with increasing
$T>T_{\rm c}(M=0)$, and remains a finite value of around 0.4 MeV at $T$ as high as 4 MeV.
At $M=$ 4 $\hbar$, whereas there is no FTBCS gap, the thermally
assisted pairing gap appears within the FTBCS1 theory at $T>$ 0,
increases with $T$ to reach a maximum at $T\sim$ 1.5 MeV, then
decreases slowly to reach the same high-$T$ limit of around 0.4 MeV
at $T\simeq$ 4 MeV.
The situation is the similar for the proton pairing
gaps, where the effect of thermally assisted
pairing correlation takes place at $M >$ 8 $\hbar$ with
the rather stable values of the gap against $T>$ 3 MeV.
\subsection{\label{backbend1}Backbending}
    \begin{figure}
        \includegraphics[width=10cm]{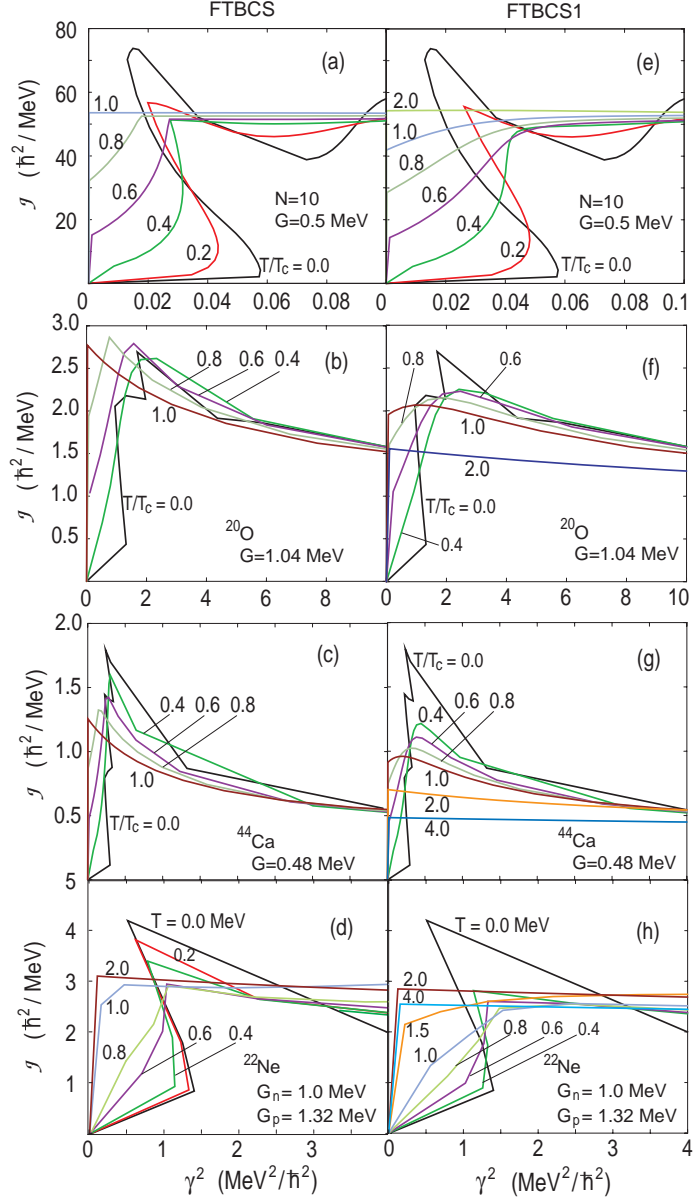}
        \caption{(Color online) Moment of inertia as a function of
        the square $\gamma^{2}$ of angular velocity $\gamma$
        obtained within the FTBCS (left) and
        FTBCS1 (right) at various $T$ for $N=10$ [(a), e)], neutrons in $^{20}$O [(b), (f)] and
        $^{44}$Ca [(c), (g)], and the whole $^{22}$Ne nucleus
        (including both proton and neutron gaps) [(d), (h)].
        \label{backbend}}
    \end{figure}
For an object that rotates
about a fixed symmetry axis, its moment of inertia ${\cal J}$
is found as the total angular momentum $M$ divided by the angular
velocity $\gamma$, i.e. ${\cal J}= M /\gamma$.
The backbending phenomenon
is most easily demonstrated by the behavior of
${\cal J}$ as a function of the square $\gamma^{2}$.
This curve first increases with
$\gamma^{2}$ up to a certain region of $\gamma^{2}$, where the
increase suddenly becomes very steep, and the curve even bends
backward. This phenomenon is understood as
the consequence of the no-crossing rule in the region of band
crossing~\cite{LL}. The SN phase transition has been suggested as
one of microscopic interpretations of backbending~\cite{MV}.

The values of the moment of inertia ${\cal J}$,
obtained at various $T$ within the schematic model as
well as realistic nuclei, is plotted in Fig.
\ref{backbend}. In the schematic model, one can
see in Figs. \ref{backbend} (a) and \ref{backbend} (e) a sharp
backbending, which takes place at
very low temperatures, $T/T_{\rm c} \leq 0.2$. As
the QNF is negligible in this temperature region,
the predictions by the FTBCS and FTBCS1 theories are almost identical.
As $T$ increases, the moment
of inertia obtained within the FTBCS changes abruptly to reach
the rigid-body value, generating a cusp, whereas, under the effect of
QNF, the value obtained
within the FTBCS1 theory gradually approaches the rigid-body value in
such a way that the cusp is smoothed out. While
no signature of backbending is seen in
the results obtained in $^{20}$O [Figs. \ref{backbend} (c) and
\ref{backbend} (f)] and $^{44}$Ca [Figs. \ref{backbend} (d) and
\ref{backbend}(g)], backbending can
be seen in $^{22}$Ne [Figs. \ref{backbend} (d) and \ref{backbend} (h)] at
$T\leq$ 0.4 MeV in agreement with the experimental data
reported in Ref. \cite{SzantoPRL42}.
\subsection{\label{correction}Corrections due to
SCQRPA and particle-number projection}
\begin{figure}
    \includegraphics[width=12cm]{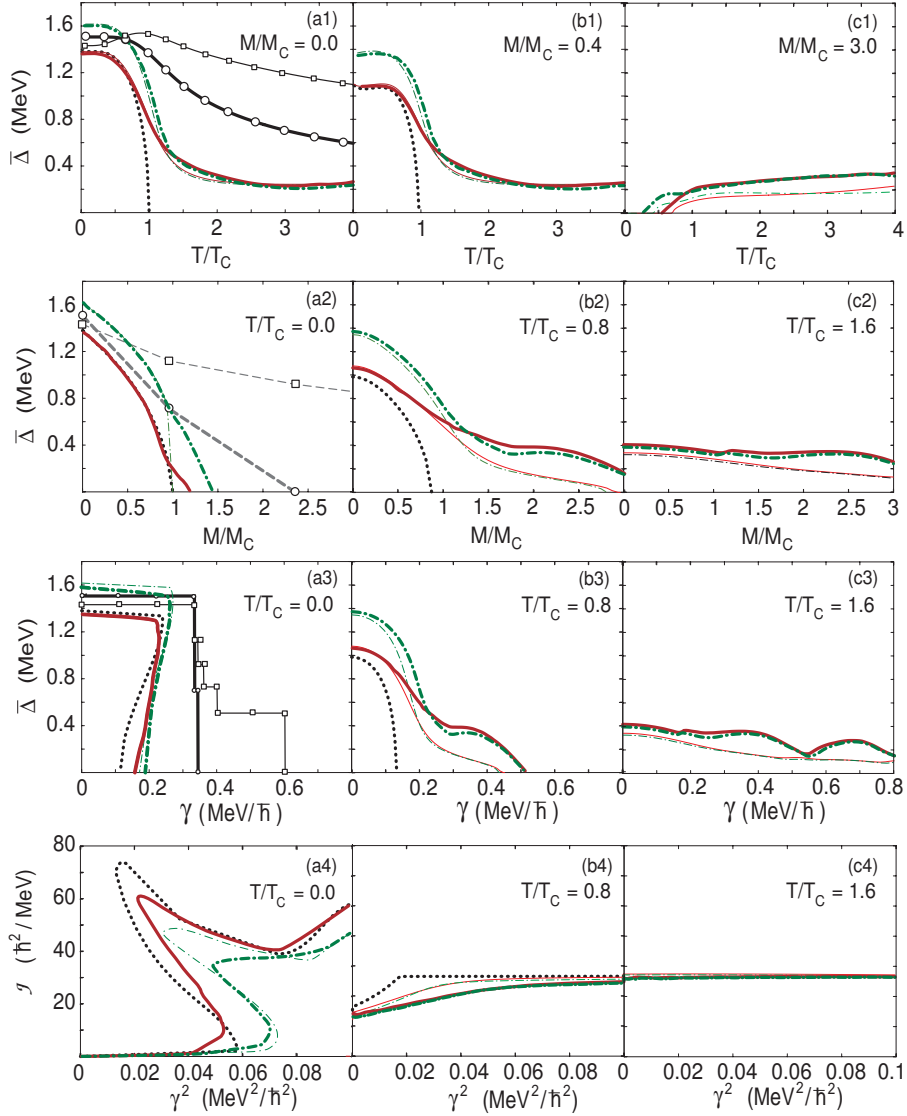}
    \caption{(Color online) Level-weighted pairing gap
    $\bar{\Delta}$ and moment of inertia ${\cal J}$ for $N =$ 10
    with $G =$ 0.5 MeV [$\varepsilon=$ 0.1 MeV in Eqs. (\ref{Momega}) and
    (\ref{gamma})].
    (a1) -- (c1): $\bar{\Delta}$ vs temperature $T$ at different
    angular momenta $M$. (a2) -- (c4): Results obtained at
    different values of $T$, namely, (a2) -- (c2):
    $\bar{\Delta}$ vs $M$;
    (a3) -- (c3): $\bar{\Delta}$ vs angular velocity $\gamma$;
    (a4) -- c4): ${\cal J}$ vs $\gamma^{2}$. The dotted, thin solid, thick
    solid, thin dash-dotted, thick dash-dotted lines are results
    obtained within the FTBCS, FTBCS1, FTBCS1+SCQRPA, FTLN1,
    FTLN1+SCQRPA, respectively.
    The solid lines with circles
    and boxes in (a1) and (a3)
    correspond to two definitions $\Delta_{\rm c}^{(1)}$ and $\Delta_{\rm
    c}^{(2)}$
    of the canonical gaps at $T=0$, respectively (See Appendix
    \ref{appendix}). In (a2) the
    dashed lines connecting the discrete values of the
    corresponding canonical gaps at $T=0$ are drawn to guide the eye.
    \label{MN10}}
\end{figure}
Shown in Fig. \ref{MN10} are the level-weighted pairing gaps and moment of
inertia, obtained within the schematic model with $N=$ 10, where predictions
offered by several approaches, namely the FTBCS, FTBCS1,
FTBCS1 + SCQRPA, FTLN1, and FTLN1 + SCQRPA, are collected.
In the same figure, the canonical gaps $\Delta_{\rm C}^{(1)}$ and
$\Delta_{\rm C}^{(2)}$ obtained at ($T\neq$ 0, $M=$ 0) [Fig. \ref{MN10} (a1)],
($T=0$, $M\neq$ 0) [Fig. \ref{MN10} (a2)], and ($T=0$, $\gamma\neq$ 0)
[Fig. \ref{MN10} (a3)], are also shown
(See Appendix \ref{appendix} for the detailed discussion of the canonical results).

As seen from Figs. \ref{MN10}, the effect due to the
SCQRPA corrections on the pairing gap increases with $M$. At $M/M_{\rm
c}\leq$ 0.8 it is rather weak,
causing only a slight enhancement of
the gap at 1.2 $<T\leq$ 2 MeV as
compared with the FTBCS1 results [Figs. \ref{MN10} (a1) and \ref{MN10}
(b1)]. However, it becomes important at $M > 1.2 M_{\rm c}$
[Figs. \ref{MN10} (c1), \ref{MN10} (a2) -- \ref{MN10}
(c2)], or $\gamma>$ 0.2 MeV$/\hbar$ (at $T\geq 0.8 T_{\rm c}$)
[Figs. \ref{MN10} (b3) and \ref{MN10}
(c3)]. In particular, the reappearance of the thermal gap at $M \geq 1.1 M_{\rm
c}$ is significantly enhanced by the SCQRPA corrections [Figs. \ref{MN10}
(c1) and \ref{MN10} (c2)]. For the moment of inertia [Figs. \ref{MN10} (a4) -- \ref{MN10}
(c4)], the SCQRPA corrections are important only at low $T$ and
$\gamma<$ 0.25 MeV$/\hbar$. At $T>$ 1 MeV, the predictions by
all the approximations for ${\cal J}$ saturate to the rigid-body value.

As compared to the predictions by the FTBCS1 and FTBCS1+SCQRPA,
the corrections due to LN-PNP are important only
at low $T$ and $M$. As the result, the gap is pushed up to be
closer to the canonical results at $T\leq T_{\rm c}$ and $M=0$ [Fig.
\ref{MN10} (a1)]. This
feature is well-known and has been discussed within the present
approach at $M=$ 0 in Ref. \cite{DangHungPRC2008}.
At $M\neq$ 0 ($\gamma\neq$ 0), the effect due to LN-PNP is noticeable in the
gaps as functions of $M$ (or $\gamma$) only at [$M\leq$ 1.2$M_{\rm c}$
($\gamma<$ 0.2 MeV$/\hbar$), $T<$ 1 MeV],
otherwise the FTLN1 (FTLN1+SCQRPA) results are hardly distinguishable from
the FTBCS1 (FTBCS1+SCQRPA) ones [Figs. \ref{MN10} (b2), \ref{MN10}
(c2), \ref{MN10} (b3), and \ref{MN10} (c3)].
Consenquently, for the moment of inertia, the LN-PNP
corrections to the FTBCS1 (FTBCS1+SCQRPA) results are important only at
($T<T_{\rm c}$, $\gamma<\gamma_{\rm c}$)
[Figs. \ref{MN10} (a4) -- \ref{MN10} (c4)]. In particular, the results
at $T=$ 0 [Fig. \ref{MN10} (a4)], where the BCS1 coincides with the BCS,
show that, backbending becomes less
pronounced within the SCQRPA and LN-SCQRPA.
For this reason, the
corrections due to LN-PNP are omitted in the results obtained for
realistic nuclei below.

\begin{figure}
        \includegraphics[width=13cm]{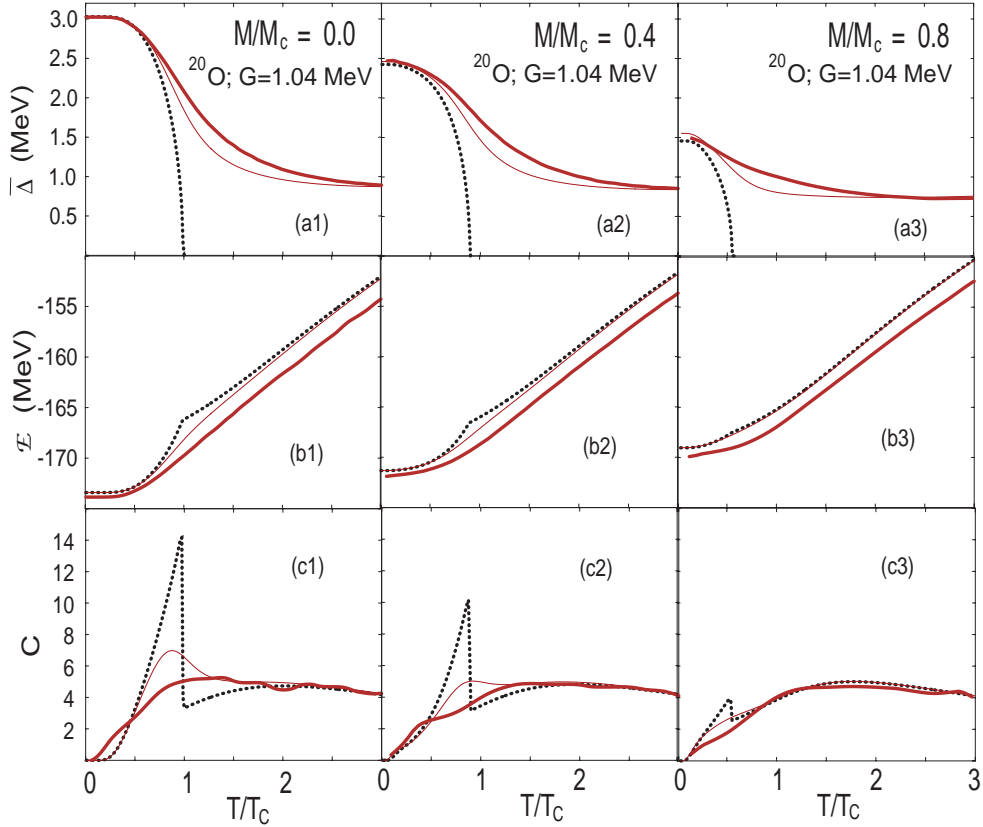}
        \caption{(Color online) Level-weighted pairing gaps
        $\bar{\Delta}$, total energies ${\cal E}$, and heat capacities
        $C$ as functions of temperature $T$ for three values of
        angular momentum $M$ obtained within
        the FTBCS (dotted lines), FTBCS1 (thin solid lines) and
        FTBCS1 + SCQRPA (thick solid lines)
        for neutrons in $^{20}$O with $G =$ 1.04 MeV ($\varepsilon=$
        0.1 MeV).
        \label{MO20}}
    \end{figure}

\begin{figure}
        \includegraphics[width=13cm]{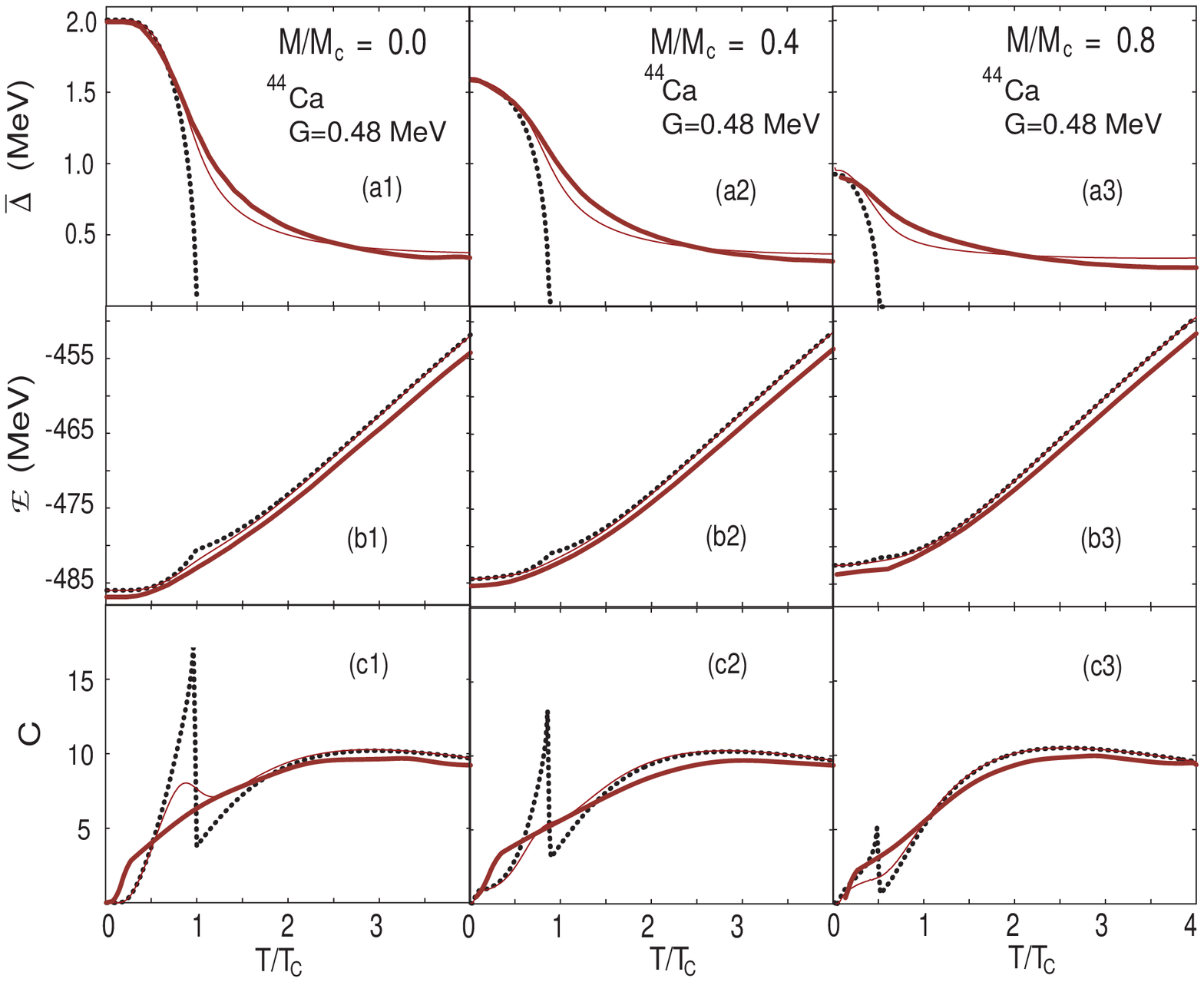}
        \caption{(Color online) Same as in Fig. \ref{MO20} but
        for neutrons in $^{44}$Ca with $G =$ 0.48 MeV
        ($\varepsilon=$ 0.1 MeV).
        \label{MCa44}}
    \end{figure}

Shown in Figs. \ref{MO20} and \ref{MCa44} are the pairing gaps, total
energies and heat capacities as functions of $T$ obtained at
$M/M_{\rm c} =$ 0, 0.4, and 0.8 within the FTBCS, FTBCS1 and
FTBCS1 + SCQRPA for realistic nuclei, $^{20}$O and $^{44}$Ca. The SCQRPA
corrections are significant for the total energy in the light nucleus,
$^{20}$O, due to the important contributions of
the screening factors \eqref{xkk} and \eqref{ykk}~\cite{HungDangPRC76,DangHungPRC2008}.
In medium $^{44}$Ca nucleus, the effect of
SCQRPA corrections on the total energy is weaker.
The corrections due to LN particle-number
projection have a similar effect as that discussed above for the
schematic model, but with much reduced magnitudes,
so they are not shown in these
figures. With increasing $M$ the pairing gap decreases. As
the result, the total energy becomes larger
but the relative effect of the SCQRPA correction does not change.
For the heat capacity, as has been reported in Ref.
~\cite{DangHungPRC2008}, the spike at $T_{\rm c}$ obtained within the FTBCS
theory, which serves as the signature of the sharp SN phase transition,
is smeared out within the FTBCS1 theory into a bump in the temperature region
around $T_{\rm c}$. With increasing $M$, this bump becomes depleted
further. Finally, the SQRPA corrections erase all the traces
of the sharp SN phase transition in the model case as
well as realistic nuclei.
\section{\label{sec4}CONCLUSIONS}
The present work extends the FTBCS1 (FTBCS1 + SCQRPA) theory to
finite angular momentum to study the pairing properties of hot
nuclei,
which rotate noncollectively about the symmetry axis.
The FTBCS1 theory includes
the quasiparticle number fluctuation whereas the FTBCS1 + SCQRPA
also takes into account the correction due
to dynamic coupling to SCQRPA vibrations.
The proposed extension is tested within the doubly degenerate
equidistant model with $N = 10$ particles as well as some
realistic (spherical and deformed) nuclei, $^{20}$O, $^{22}$Ne, and
$^{44}$Ca. The numerical calculations were carried out within the FTBCS,
FTBCS1, and FTBCS1 + SCQRPA for the pairing gap,
total energy, and heat capacity as functions of
temperature $T$, total angular momentum $M$, and angular velocity
$\gamma$. The corrections due to the Lipkin-Nogami particle-number
projection are also discussed.
The analysis of the numerical results- allows us to draw the
following conclusions:

1. The proposed approach is able to reproduce the
effect of thermally assisted pairing correlation that takes place in
the schematic model within the FTBCS
theory, according to which
a finite pairing gap can reappear within a given temperature interval, $T_1 <
T < T_2$ ($T_1 > 0$), while it is zero beyond this interval. However,
this phenomenon does not show up in realistic nuclei under consideration.

2. Under the effect of QNF, the paring gaps obtained within the
FTBCS1 at different values $M$ of angular momentum
decrease monotonously as $T$ increases, and
do not collapse even at hight $T$ in the schematic model as well as realistic
nuclei. The effect of thermally assisted pairing correlation is seen
in all the cases, but in such a way that the pairing gap
reappears at a given $T_1 >$ 0 and remains finite at $T > T_1$, in
qualitative agreement with the canonical results of Ref. \cite{FrauendorfPRB68}.

3. The backbending of the moment of inertia is found in the schematic
model and in $^{22}$Ne in the low temperature region, whereas
it is washed out with increasing temperature. This effect does not
occur in $^{20}$O and $^{44}$Ca, in consistent with existing
experimental data and results of other theoretical approaches.

4.
The effect caused by the
corrections due to the dynamic coupling to SCQRPA vibrations
on the pairing gaps, total energies, and heat capacities
is found to be significant in the region around
the critical temperature $T_{\rm c}$
of the SN phase transition and/or at large angular
momentum $M$ (or angular velocity $\gamma$).
It is larger in lighter systems. As the result,
all the signatures of the sharp SN phase
transition are smoothed out in both schematic model and realistic nuclei.
The SCQRPA corrections also significantly enhance the reappearance of the thermal
gap at finite angular momentum. On the other hand, the effect caused by the
corrections due to PNP is important only at
temperatures below $T_{\rm c}$, and at quite low angular momentum. In
particular, it makes backbending less pronoucned at $T=$ 0.

Still the fluctuations due to violation of angular-momentum
conservation are not implemented in the present extension. We
hope that further studies in this direction will shed light
on this issue.
\begin{acknowledgments}
The authors thank L.G. Moretto (Berkeley) for suggestions, which led
to the present study. Fruitful discussions with S. Frauendorf
(Notre Dame), and P. Schuck (Orsay) are
acknowledged. NQH is a RIKEN Asian Program Associate.
The numerical calculations were carried out using the {\scriptsize FORTRAN} IMSL
Library by Visual Numerics on the RIKEN Super Combined Cluster
(RSCC) system.
\end{acknowledgments}
\appendix
\section{On the comparison with canonical results}
\label{appendix}
\begin{figure}
    \includegraphics[width=12cm]{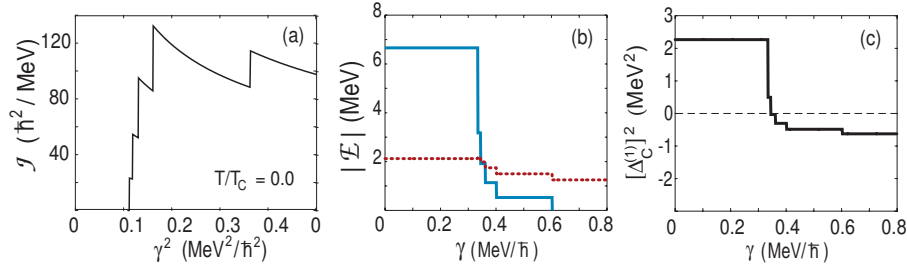}
 \caption{(Color online) (a) Canonical moment of inertia vs
 $\gamma^{2}$; (b): Absolute values
 $|\langle{\cal E}\rangle_{\rm C}-{\cal
 E}_{\rm m.f.}|$ (solid line) and $|{\cal E}_{\rm unc.}|$ (dotted line)
 vs $\gamma$; (c): $[\Delta_{\rm C}^{(1)}]^{2}$ vs
 $\gamma$ for the schematic model with $N=$ 10 and $G=$
 0.5 MeV at $T=$ 0.
 \label{append}}
    \end{figure}
It has been shown in Sec. \ref{model} that
the FTBCS1 (FTBCS1+SCQRPA) produces results
in qualitative agreement with the canonical
ones of Ref. \cite{FrauendorfPRB68}, in particular, the reappearance
of the thermal gap at $M\neq$ 0. However, it is important to make
clear the difference between the predictions by BCS-based approaches and the
canonical results.
As a matter of fact, the $z$-projection $M$
of the total angular momentum  within the FTBCS (FTBCS1) approach is
temperature-independent.
At $T$ varies, by solving the FTBCS
(FTBCS1) equations,
the angular velocity $\gamma$ is defined as a
Lagrangian multiplier so that
$M$, being the thermal average of the total angular
momentum within the grand canonical ensemble (\ref{average}),
remains unchanged. In this way, within the FTBCS (FTBCS1), the
angular velocity $\gamma$ varies with $T$, whereas
$M$ does not. Similar
to that for choosing the chemical potential
$\lambda$ to preserve the (grand-canonical ensemble)
average particle-number $N$, this contraint is physically reasonable when the
total angular momentum is conserved as in the noncollective rotation
of spherical systems or rotation of axially symmetric systems about
the symmetry axis, as has been discussed in Sec.
\ref{subsec1}.

On the contrary, the canonical results in Ref.
\cite{FrauendorfPRB68} are obtained by embedding the eigenvalues
$E_{\nu,i}(\gamma)=E_{\nu}-\gamma M_{\nu,i}$ in the canonical ensemble with
the partition function
\begin{equation}
Z(\beta,\gamma)=\sum_{\nu,i}e^{-\beta
E_{\nu,i}(\gamma)}~.
\label{Zc}
\end{equation}
Here $E_{\nu}$ denote the eigenvalues of the $\nu$th state
with seniority $\nu$ at $\gamma=$ 0, whereas
$M_{\nu,i}$ are the z-projections of angular momenta of $\nu$
nucleons. While the eigenvalues $E_{\nu}$ are obtained by separately
diagonalizing the pairing Hamiltonian $H_{P}$ in Eq. (\ref{Ha1}),
the rotational part $\Phi_{\nu}=\sum_{i}\exp(\beta\gamma M_{\nu,i})$
of the partition function $Z(\beta,\gamma)$ is calculated following
Ref. \cite{Kuzmenko}. The resulting canonical average value $\langle
M(\beta,\gamma)\rangle_{\rm C}=
\beta Z(\beta,\gamma)^{-1}\partial{Z(\beta,\gamma)}/\partial{\gamma}$
of angular momentum, therefore, varies with $T$.
On the other hand, the angular velocity
$\gamma$ just plays the role of an
independent parameter, therefore, does not depend on $T$.
By the same reason, each
canonical average value $\langle M(\beta,\gamma)\rangle_{\rm C}$ corresponds to a
single value of $\gamma$, i.e. the
canonical moment of inertia ${\cal J}_{\rm C}$ undergoes no
backbending, as shown in Fig. \ref{append} (a).

Because of this principal difference,
a quantitative comparison between the FTBCS (FTBCS1) results,
and the canonical ones as functions of $M$ (or $\gamma$) at $T\neq$ 0
unfortunately turns out to be impossible.
To establish a meaningful correspondence,
one needs to know the exact eigenvalues
of the ground state as well as all excited states of the pairing
problem described by Hamiltonian
(\ref{Ha}) so that,
by embedding the eigenvalues in the grand canonical ensemble,
$\gamma$ becomes a function of $T$ in such a way to keep
$\langle M(\beta,\gamma)\rangle_{\rm C}$ always equal to $M$. To our
knowledge, this problem still remains unsolved. One may also try to
estimate the results within the microcanonical ensemble. However,
here one faces a problem of extracting the nuclear temperature,
which is rather ambiguous at low level density (small $N$) within
the schematic model under consideration~\cite{Sumaryada,DangHung},
whereas the extension
of exact solution of the pairing problem to $T\neq$ 0 is unpracticable
at $N\geq$ 16.

Therefore, in the present paper, we can only compare
the predictions of our approach with the canonical results
as functions of temperature $T$ at $M=$ 0,
or as functions of $M$ (or angular
velocity $\gamma$) at $T=$ 0.
For this purpose, and given several definitions of the ``effective''
gaps existing in literature, we choose to employ in the present paper two definitions
of the canonical gaps, $\Delta_{\rm C}^{(1)}$ and $\Delta_{\rm C}^{(2)}$.
They should be understood as effective ones since a gap
{\it per se}, which is a mean-field concept,
does not exist in the exact solutions of the pairing problem.

The canonical gap $\Delta_{\rm C}^{(1)}$
is defined from the pairing energy ${\cal E}_{\rm pair}$
of the system as
\begin{equation}
    [\Delta_{C}^{(1)}]^{2}={-G{\cal E}_{\rm pair}}~,\hspace{3mm}
    {\cal E}_{\rm pair}\equiv\langle{\cal
    E}\rangle_{\rm C}-{\cal
    E}_{\rm m.f.}-{\cal E}_{\rm unc.}~, \hspace{3mm} {\cal
    E}_{\rm m.f.}\equiv
    2\sum_{k}\epsilon_{k}f_{k}~,\hspace{3mm}
    {\cal E}_{\rm unc.}\equiv-{G}\sum_{k}f_{k}^{2}~.
    \label{gapC1}
    \end{equation}
Here $\langle{\cal E}\rangle_{\rm C}$ is the total energy
within the canonical ensemble with the partition function
$Z(\beta,\gamma)$ given by Eq. (\ref{Zc})
of a system rotating at angular velocity $\gamma$, or
with the partition function $Z(\beta, 0)$ at $M=$ 0.
The term ${\cal E}_{\rm m.f.}$
denotes the energy of the single-particle motion described by the first term
at the right-hand side of the pairing Hamiltonian $H_{P}$ in Eq. (\ref{Ha1}).
Functions $f_{k}$ are
occupation numbers of $k$th orbitals within the canonical ensemble.
The energy ${\cal E}_{\rm m.f.}$ becomes that of the mean-field
once the single-particle occupation
numbers $f_{k}$ are replaced with those describing the
Fermi-Dirac distributions of independent particles. The
energy ${\cal E}_{\rm unc.}$
comes from the uncorrelated single-particle
configurations caused by the pairing interaction in Hamiltonian
(\ref{Ha1}). Therefore, by subtracting the term
${\cal E}_{\rm m.f.}+{\cal E}_{\rm unc.}$
from the total energy $\langle{\cal E}\rangle_{\rm C}$, one obtains the
result that corresponds to the energy
due to pure pairing correlations. The definition (\ref{gapC1})
is very similar to that given in Ref. \cite{Delft}. It is, however,
different from the canonical gap $\Delta_{\rm C}^{(2)}$, which
is used in Refs. \cite{FrauendorfPRB68}. The latter is defined as
\begin{equation}
    [\Delta_{\rm C}^{(2)}]^{2}={-G\big[\langle{\cal E}\rangle_{\rm C}-
    \langle{\cal E}(G=0)\rangle_{\rm C}\big]}~,
    \label{gapC2}
    \end{equation}
where $\langle{\cal E}(G=0)\rangle_{C}$
is the total canonical energy $\langle{\cal E}\rangle_{\rm C}$
at $G=$ 0.

The canonical gaps  $\Delta_{\rm C}^{(1)}$ and $\Delta_{\rm C}^{(2)}$ are
shown in Figs. \ref{MN10} (a1), \ref{MN10} (a2), and \ref{MN10} (a3)
as functions of temperature $T$ (at
$M=$ 0), angular momentum $M$ (at $T=$ 0),
and angular velocity $\gamma$ (at $T=$ 0), respectively.
It is seen from these figures that the difference between
the two canonical gaps $\Delta_{\rm C}^{(1)}$ and $\Delta_{\rm
C}^{(2)}$ is rather significant at large $T$ for $M=$ 0, and at large
$M$ (or $\gamma$) for $T=$ 0. The reason is rather simple since the definition
(\ref{gapC1}) of $\Delta_{\rm C}^{(1)}$ is rather similar to that for
the BCS gap.  As a matter of fact, by replacing the canonical
single-particle
occupation numbers $f_{k}$ with the Bogoliubov's coefficients
$v_{k}^{2}$, and the total energy
$\langle{\cal E}\rangle_{\rm C}$ with that obtained within the BCS
theory, the gap $\Delta_{\rm C}^{(1)}$
reduces to the usual BCS gap. Meanwhile, by doing so with $\Delta_{\rm
C}^{(2)}$, the energy $\langle{\cal E}(G=0)\rangle_{\rm C}$
just reduces to the Hartree-Fock energy, leaving the uncorrelated
energy ${\cal E}_{\rm unc.}$ out of the definition.
Consequently, as functions of $T$, the gaps predicted by the BCS-based
approaches under consideration agree better with the
canonical gap $\Delta_{\rm C}^{(1)}$ than with
$\Delta_{\rm C}^{(2)}$ [Fig.
\ref{MN10} (a1)].

As functions of angular velocity $\gamma$,
both the squared values (\ref{gapC1}) and
(\ref{gapC2}) of the canonical gaps undergo a stepwise
decrease with increasing $\gamma$. The step occurs whenever the state
of the lowest energy changes from $\nu-2$ to $\nu$, causing a stepwise
increase of $\langle M(\beta,\gamma)\rangle_{\rm C}$~\cite{FrauendorfPRB68}.
Therefore, for $N=$ 10, the pairs are gradually broken in 5 steps with
a corresponding stepwise increase of seniority $\nu$ from 0 to 10 by two
units in each step. However, Fig. \ref{append} (b)
shows that the absolute value of the
uncorrelated energy ${\cal E}_{\rm unc.}$, which enters
in the definition (\ref{gapC1}) of the gap $\Delta_{\rm
C}^{(1)}$, becomes larger than that of the difference ${\cal E}_{\rm C}-{\cal
E}_{\rm m.f.}$ already at the second step, leading to
$[\Delta_{\rm C}^{(1)}]^{2}<$ 0 [Fig. \ref{append} (c)], i.e. an
imaginary value for $\Delta_{\rm C}^{(1)}$. As the result, instead of
collapsing as $\Delta_{\rm C}^{(2)}$ in 5 steps at a rather large value of
$M$ (or $\gamma$), the canonical gap
$\Delta_{\rm C}^{(1)}$ collapses in two steps at a value of $M$
(or $\gamma$) much closer to $M_{\rm c}$ (or
$\gamma_{\rm c}$) for the BCS gap [Figs. \ref{MN10} (a2) and \ref{MN10} (a3)].
Once again, this makes
the gaps predicted by the BCS-based
approaches as functions of $M$ (or $\gamma$) agree better with the
canonical gap $\Delta_{\rm C}^{(1)}$, rather than with
$\Delta_{\rm C}^{(2)}$ [Figs.
\ref{MN10} (a2) and \ref{MN10} (a3)].


\end{document}